\documentclass[journal]{IEEEtran}

\ifCLASSINFOpdf
\else
\fi

\usepackage{etex}
\usepackage{algorithmicx}
\usepackage{amsmath,amssymb,amsthm} 
\usepackage{bbding}
\usepackage[noadjust]{cite}
\usepackage[final]{pdfpages}
\usepackage{array}
\usepackage{graphicx}
\usepackage{xcolor,colortbl}
\usepackage{epstopdf}
\usepackage{setspace}
\usepackage{ctable}
\usepackage{enumerate}
\usepackage{ragged2e}
\usepackage{amssymb}
\usepackage{multirow}
\usepackage{tabularx,tabulary}
\usepackage{nicefrac}		%
\usepackage[caption=false,font=footnotesize,labelformat=simple]{subfig} %
\usepackage[flushleft]{threeparttable}
\usepackage[bookmarks=false]{hyperref}
\usepackage{rotating}
\usepackage{tabularx,ragged2e,booktabs}
\newcolumntype{C}[1]{>{\Centering}m{#1}}

\usepackage{mathtools}
\usepackage{balance}

\usepackage{authblk}
\usepackage{verbatim}

\usepackage{xcolor}
\usepackage{tcolorbox}
\newtcbox{\highlight}[0]{boxsep=0pt,left=0pt,top=0pt,bottom=0pt,right=0pt,boxrule=0pt,arc=0pt,auto outer arc,colback=green,width=15cm}

\definecolor{dark-blue}{RGB}{0,70,127}

\usepackage{pifont}
\usepackage{framed}
\usepackage{soul}
\usepackage{booktabs}
\usepackage{multirow}

\definecolor{lightgreen}{RGB}{153,237,137}
\definecolor{lightgray}{gray}{0.9}

\usepackage[ruled,vlined]{algorithm2e}
\SetKwRepeat{Do}{do}{while}

\newcommand{\revision}[1]{\textcolor{black}{#1}}

\usepackage[export]{adjustbox}
\usepackage{pdfpages}

\definecolor{Gray}{gray}{0.9}

\begin{document}
\title{A Survey of Cybersecurity of  Digital Manufacturing}

\author{Priyanka Mahesh, Akash Tiwari, Chenglu Jin, Panganamala R. Kumar,~\IEEEmembership{Fellow,~IEEE}, A.~L.~Narasimha~Reddy,~\IEEEmembership{Fellow,~IEEE}, Satish T.S. Bukkapatanam, Nikhil Gupta, and Ramesh Karri, ~\IEEEmembership{Fellow,~IEEE}%
\thanks{P. Mahesh, C. Jin, N. Gupta, and R. Karri are with New York University, Brooklyn,
NY, 11201 USA. e-mail: \{pm2929, chenglu.jin, ngupta, rkarri\}@nyu.edu}%
\thanks{A. Tiwari, P. R. Kumar, A. L. N. Reddy, and S. T. S. Bukkapatanam are with Texas A\&M University, College Station, TX, 77843 USA. e-mail: \{akash.tiwari, prk, reddy, satish\}@tamu.edu }%
}

\maketitle

\begin{abstract}
\revision{The Industry 4.0 concept promotes a digital manufacturing (DM) paradigm that can enhance quality and productivity, that reduces inventory and the lead-time for delivering custom, batch-of-one products based on achieving convergence of Additive, Subtractive, and Hybrid manufacturing machines, Automation and Robotic Systems,  Sensors, Computing, and Communication Networks, Artificial  Intelligence, and Big Data. A DM system consists of embedded electronics,  sensors,  actuators,  control software, and inter-connectivity to enable the machines and the components within them to exchange data with other machines, components therein, the plant operators, the inventory managers, and customers. This paper presents the cybersecurity risks in the emerging DM context, assesses the impact on manufacturing, and identifies approaches to secure DM. }
\end{abstract}


\IEEEpeerreviewmaketitle

\section{Introduction}

\IEEEPARstart{D}{igitalization} of manufacturing aided by advances in sensors, artificial intelligence, robotics, and networking technology, is revolutionizing the traditional manufacturing industry by rethinking manufacturing as a service. 
Concurrently, there is a shift in demand from high volume manufacturing to batches-of-one, custom manufacturing of products~\cite{zhong2017intelligent}. 
While the large manufacturing enterprises can reallocate resources and transform themselves to seize these opportunities, the medium and small scale enterprises (MSEs) with limited resources need to become federated and proactively deal with digitalization. %
Many MSEs essentially consist of general-purpose machines that give them the flexibility to execute a variety of process plans and workflows to create one-off products with complex shapes, textures, properties, and functionalities. %
One way the MSEs can stay relevant in the next generation digital manufacturing (DM) environment is to become fully inter-connected with other MSEs by using the digital thread and becoming part of a larger, cyber-manufacturing business network~\cite{Iquebal2018towards}. %
This allows the MSEs to make their resources visible to the market and continue to receive work orders\footnote{MSEs serve as suppliers to OEMs and other parts of the manufacturing supply networks.}. Digitization will also enhance compliance with the larger industry and customers in terms of technology standards and practices, and access resources and services available through the inter-connected digital supply chain (DSN) network.

In the emerging DM, timeliness of information is important for lean production, as well as quality and productivity assurance. Digitization creates communication channels across vendors and OEMs on one hand and between the various machines and processes inside an MSE on the other.  
DM requires the integration of cyber (computing and communications) resources with the physical resources in the manufacturing process and supply chain. Continuous streaming of data from sensors at various locations in the manufacturing plant (e.g., individual machines and the network of machines) informs the data-driven decision making that guides design modifications, calibrates manufacturing methods, and programs the robot tasks and paths that they navigate the manufacturing floor. Securing such a distributed and connected cyber-physical system against cyberattacks requires developing novel approaches that are tailored to the threats faced by such systems. %
The cyberattacks can range from sabotage of product quality and intellectual property theft to ransomware. The attack surface, threat vectors, and solutions need to be analyzed to enable a secure, resilient, and scalable next generation DM.
\begin{figure*}[ht]
    \centering
    \includegraphics[width=\textwidth]{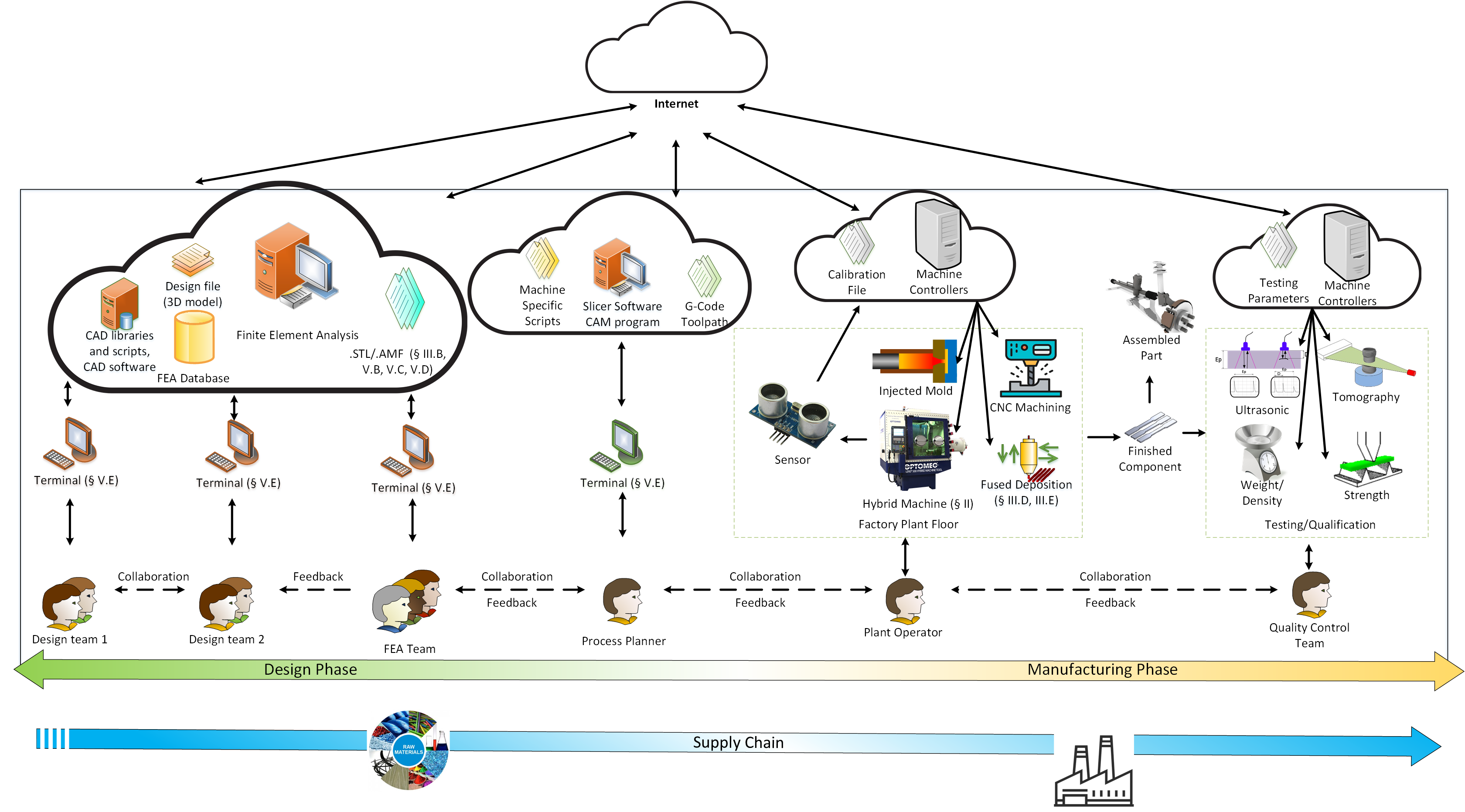}
    \caption{\revision{A representative process workflow in digital manufacturing (DM) systems. The workflow is broadly divided into design and manufacturing phases. The design teams, Finite Element Analyst (FEA) teams, and Process Planners come up with designs and manufacturing processes. The plant operators operate DM machines, and the finished components are tested by quality control teams using various testing methods. In future DM systems, the workflow, including design and manufacturing phases, will run in clouds since they provide flexibility, reliability, and connectivity. Also, this new paradigm of DM systems come with security concerns that we will address in this paper. In this Figure, we explicitly point to (sub)sections in the paper where the topic is addressed.}}
    \label{fig:supply_chain}
\end{figure*}

Traditionally, manufacturing plants have been siloed and naturally create air gaps making them secure~\cite{tuptuk2018security}. On one hand, DM exploits the information from the various sensors and devices to streamline the process and material flow. On the other hand, the distributed and collaborative nature of DM exposes it to risks that come with the connectivity required to implement DM. A typical DM process workflow is illustrated in  Figure~\ref{fig:supply_chain}. A large part of the process before the actual manufacturing step is completely digital and relies on  computational resources and computer networks for design, simulation, and programming the controllers of the manufacturing machines. The DM system may consist of additive, subtractive, and hybrid manufacturing machines. This process flow requires connectivity throughout the process chain.  %
However, connectivity poses a security risk, which needs to be addressed by traditional and novel cybersecurity solutions that apply to various steps of the process flow. %
This paper \revision{presents the hybrid machine tool as an archetype for DM}, analyzes the cybersecurity risks, develops an attack taxonomy and proposes novel solutions for the DM cyber-physical system.

This paper is organized as follows:  \revision{Section~\ref{sec:hybrid} will present a  hybrid manufacturing cell, a building block of DM, and uses it to discuss vulnerabilities. A taxonomy of threats for DM and attack case studies are discussed in Section \ref{sec:threats}. A survey of existing taxonomies in digital manufacturing systems is presented in Section~\ref{sec:comparison}.} Section \ref{sec:defense} will demonstrate how novel manufacturing-unique defenses can mitigate the attacks. Section \ref{sec:conclusion} discusses lessons learned from state-of-the-art in DM security and  research challenges.

\section{Hybrid Machine Tool: A DM Basic Block}\label{sec:hybrid}

\revision{Hybrid machine tools are excellent archetypes of a DM building block. They make for a case to explain how traditional manufacturing is transforming into DM. This resulting transformation however creates additional attack vectors for DM. The case of a hybrid machine tool therefore allows to identify, analyze and address the vulnerabilities from these attack vectors before the widespread adoption of DM.}

\revision{The most common configurations of hybrid machine tools combine additive and subtractive manufacturing processes on the same platform \cite{Kerfees1} so that process chains spread across multiple machines (possibly located at different enterprises) can be carried out within a single machine. This is especially beneficial for the fabrication of custom components, as it results in reduced setup times, material costs and error in handling. Hybrid machines have been increasingly considered in the industry for re-manufacturing and repair of high value components and in manufacturing parts that require complex process chains. Pipe casings for offshore oil extraction have several features (e.g., Fins and Spiral coatings) on the surface critical. The use of a hybrid machine for such a part was proven to reduce material cost by $\sim$97.2\% in addition to the tooling cost \cite{Yamazaki}. A hybrid machine can customize implants by milling the implant-abutment interface followed by printing the abutment custom designed for a patient \cite{Daniel}, create novel injection molds with improved cooling performance over traditional fabrication methods \cite{Masakazu}, and enable surface patching in mold and die repair \cite{Lan} and turbine blade repair \cite{Zelinski}. }

\revision{More pertinently, hybrid machine tools are  oftentimes integrated with state-of-the-art digital information technology (IT) systems (e.g., software and data warehouse) and operational technology (OT) systems (e.g., sensors and communication channels) to work in tandem to produce the desired part \cite{SHM}. Integration of digital technologies provides the connectivity and computational infrastructure for enabling a hybrid machine tools to be part of a DM network.}
Connectivity includes the feedback loops within the machines based on the machine state, and feedback loops based on the observations of the process from an observer external to the machine. It also refers to the communication channels among the manufacturing resources within the manufacturing cell. The computational infrastructure supports data collection, storage, analysis, and decision making elements of manufacturing. While connectivity and computational infrastructure improve the utilization of the manufacturing resources, they can be attack vectors for internal and external adversaries. 

\revision{Due to the use of IT/OT technologies in these hybrid machine tools, much of the threats these systems face are similar to those of the conventional IT/OT technologies. However, the sabotaging effect of these threats pose direct safety and productivity challenges to the manufacturing enterprise. For example, traditional cyber-security attacks on legacy systems connected to IT/OT technologies in the recent past have resulted in machine downtime, idle time and reduced reliability of the system causing massive monetary losses to the enterprise. }

Vulnerable nodes in the supporting infrastructures must be identified and secured to realize the economic and efficiency benefits of DM. In the following sections we describe key components of our use-case hybrid machine and the possible vulnerabilities in the context of DM. %
\revision{The discussion on the identified vulnerabilities of the hybrid machine serve as a motivation for the development of taxonomy and the solutions to the vulnerabilities for DM in the rest of the paper.}

\begin{figure*}[t!]
    \centering
    \subfloat[\label{fig:hybridmachinetool_elements1}]{%
       \includegraphics[width=\columnwidth]{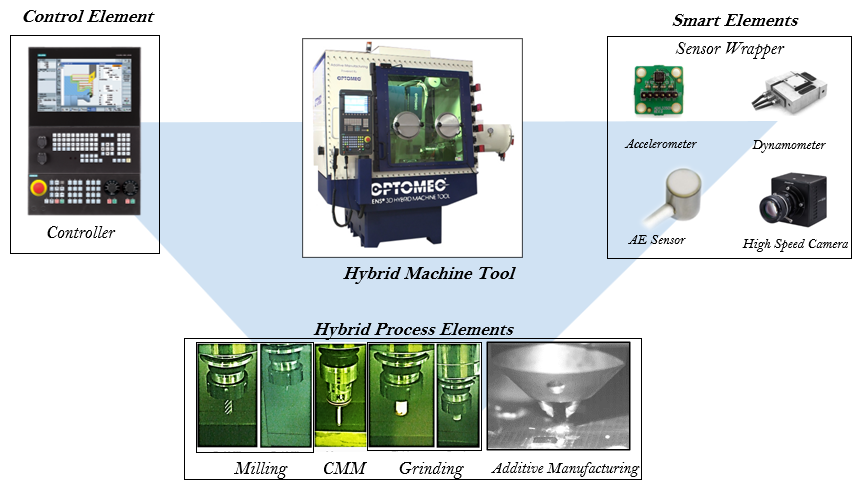} %
     }
      \subfloat[\label{fig:hybridmachinetool_elements2}]{%
       \includegraphics[width=\columnwidth]{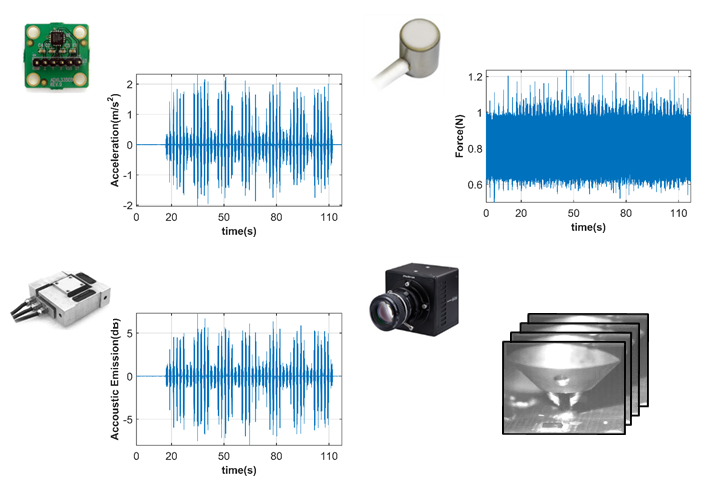}
     }
    \caption{(a) Texas A\&M University's Smart Hybrid Machine Tool with its constituent elements. (b) Data streams collected from the sensor wrapper of the smart element~\cite{BotchaMfgLett20}}
    \label{fig:hybridmachinetool_elements}
\end{figure*}

\subsection{Transforming a hybrid machine (HM) as part of a DM ecosystem}
\revision{Figure \ref{fig:hybridmachinetool_elements1} illustrates a HM located at Texas A\&M University.} It consists of three key elements – the hybrid process element, the controller, and the smart element \cite{BotchaMfgLett20}. The hybrid process elements include the milling tools, the coordinate measuring touch probe, grinding tools, and the laser engineered net-shaping process that employs a directed energy deposition printing head. These tools support consecutively running the additive and subtractive manufacturing operations within a process cycle. The control element allows the user to interface with the hybrid process element and the execution of process cycles. It acts as an internal observer that gathers the internal state of the machine (e.g., position, feed rate, laser power, and spindle speed) and sends actuation signals based on the instructions specified by the operator. The smart elements include sensors with supporting hardware. Hardware and software that enable data acquisition from the sensors are termed the sensor wrapper \cite{10.1115/MSEC2018-6726,IQUEBAL202029}. The sensor wrapper implementation is composed of high-resolution sensors (here, accelerometer, acoustic emission sensor, dynamometer, and a high-speed camera), data acquisition system, signal conditioning elements such as filters and amplifiers, and human machine interface. The sensor signals allow the process states to be estimated during a process cycle for feedback control \cite{Rao2014real} as well as for providing observations from the perspective of an external observer (e.g., the operator) \cite{palanna2003model}. The three elements of the HM work in harmony to enable refined control over the process. Such harmony is possible due to the coordination among process hardware and IoT devices in the computing and the communication channels.

\revision{The very capabilities of a HM tool that allows fabrication of parts with complex geometries and functionalities (as it combines multiple manufacturing processes into a single platform), as noted earlier, create complexities in the process cycles and allow for faults to creep into the process.} While process faults are inevitable for any complex system, one needs to execute corrective measures to mitigate the effects of these faults. Monitoring the process as an external observer is therefore essential in operating the hybrid machine tool. The hybrid elements can allow the operator to take corrective actions when a fault is observed. For example, a defect created in the part during the additive manufacturing cycle can be undone by executing a subtractive cycle over the layer with the defect before resuming the additive cycle. 
Taking corrective measures after a fault occurs leads to loss in manufacturing lead time and the physical resources. The smart elements can intervene to save time and resources by informing the operator about an imminent fault. This is possible by using the information that the sensor wrapper collects. Figure \ref{fig:hybridmachinetool_elements2} illustrates the time synchronized data stream for an additive manufacturing cycle collected over 120 seconds. %
The data stream for the force signals are densely packed, therefore an adjacent plot represents the force plot for a 0.05 second window. The information generated from the sensor wrapper is voluminous. The data streams from acoustic emission, the accelerometer, and the force transducers,  over a 120 second period generate 89.5 MB, 44.7 MB and 8.92 MB of data, respectively. The High-speed camera generates 110 GB streaming image data over the 120 second period.

\begin{figure}[t!]
    \centering
    \includegraphics[width=\columnwidth]{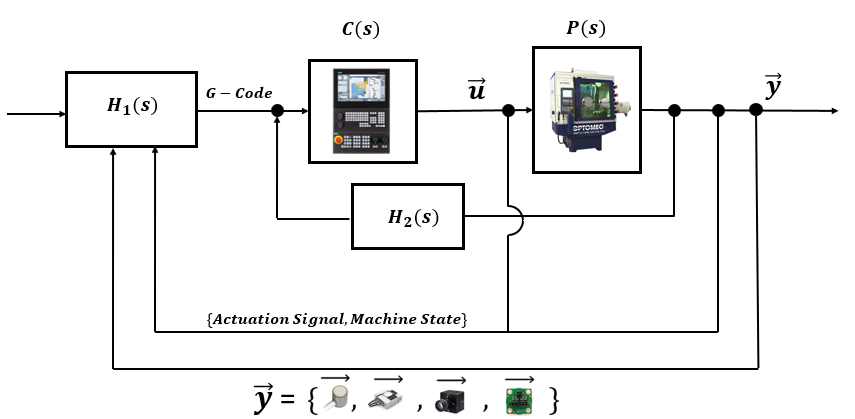}
    \caption{Closed Loop control block diagram for the Hybrid Machine Tool.}
    \label{fig:block_diagram}
\end{figure}
The controller (internal observer) observes and controls the HM tool based on the machine state. The external observer however, observes the process and takes corrective measures. This establishes two feedback loops. The controller sends actuation signals to the HM tool based on instructions within the G-code (subject to change based on the external observations of the process) that is sent by the operator. The G-code file contains high-level instructions meant to be executed on the HM. The operator may observe the information stream and take corrective measures by sending new instructions when the information stream resembles the nascent stages of an imminent fault, thereby overcoming the fault altogether. This is illustrated in Figure \ref{fig:block_diagram} as a closed loop controller. 

The refined control over the process is thus achieved by a feedback control that is based on both – information on the machine state and information about the process. The feedback control entails collecting, processing, and analyzing voluminous information to derive inferences about the process in real time. This requires computing on large amounts of information in a timely manner and may resort to AI methods to process the information. This makes the need for computing infrastructure apparent. Factors influencing the computing infrastructure include, the environment where computing happens, latency of the computation, the type of data, and the amount of data. 

In online quality control where the corrective and prognostic measures are to be taken, information from the sensor is processed in real time to infer about the state of the process and therefore, data storage and computing resources must be in the vicinity of the process to avoid latency. Another situation for online quality control is where latency of the calculation is not an issue, but there are no computational resources on the shop floor. Then, the computational services offered by cloud platforms are leveraged. For offline quality control, where a defect in the part is identified later, the investigator may use data collected during the process to identify process faults --missed by online quality control-- that may have led to a defect. %

\begin{figure}[b!]
    \centering
    \includegraphics[width=\columnwidth]{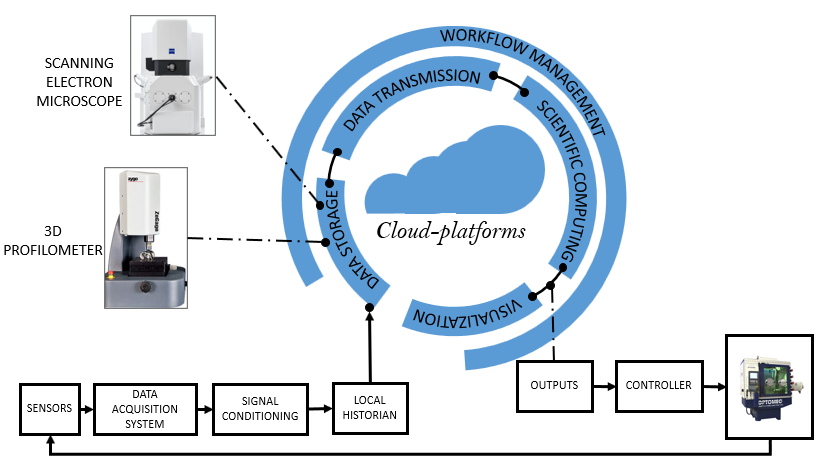}
    \caption{Cloud computing platform for a Hybrid Manufacturing (HM) Cell.}
    \label{fig:computing_infras}
\end{figure}

Thus, the computing infrastructure is dictated by the requirements of the manufacturing cell. Data storage, computations, and transmission of the calculations to the destination are essential to establish the closed loop control. Since manufacturing shop-floors may be limited in their capacity to cater to such requirements efficiently, cloud computing infrastructure  could be economical and efficient. Cloud computing infrastructure is mature and reliable for application in the hybrid manufacturing cells. Cloud service providers (e.g., Amazon Web Services and Rackspace) have integrated the elements of storage, computation and communication. Amazon provides storage services (the Elastic Block Store) and hosts well-known software (R, Matlab, Mathematica) as Virtual Machines (VMs). All computations can be visualized on the cloud VMs with software like Tableau. The workflow in the cloud can be orchestrated by scientific workflow software such as Kepler. 

Figure \ref{fig:computing_infras} illustrates the cloud as central to online and offline quality control for the HM cells. Signals collected by the sensors from the plant are stored in a local historian and uploaded to the cloud for storage. From this point, the scientific workflow management software handles the flow of data. The computing VM is activated to receive and analyze the data, and to calculate new control outputs, which are downloaded onto the controller, closing the loop. For offline quality control, scanning electron microscopes and 3D profilometers in the HM cell inspect the part. These instruments download process-related data streams from the cloud and identify anomalies in the process to explain defects in the part.

\subsection{Vulnerabilities in a HM Tool}
\begin{figure}[t!]
    \centering
    \includegraphics[width=\columnwidth]{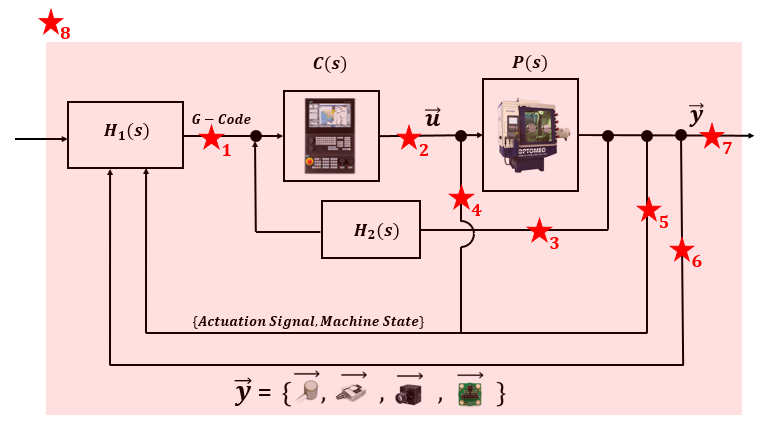}
    \caption{Vulnerable nodes in a HM Tool. The vulnerable nodes are identified by a red star, indexed by a subscript.}
    \label{fig:vnodes_hybrid}
\end{figure}

Although the HM tool is only one of the multiple resources of a DM process workflow, this critical resource has multiple vulnerable nodes. \revision{\cite{r_cardenas} identified possible attacks on cyber-physical system and discussed theoretical formulation for the attacks to be addressed and the requirements of a secure cyber-physical system using extant theory in controls, information security and network security. The issues identified in \cite{r_cardenas}  were however generalized for a cyber-physical system.} Likewise, specific to securing a DM system, Figure \ref{fig:vnodes_hybrid} identifies eight vulnerable nodes in the closed loop control diagram shown in Figure \ref{fig:block_diagram}.  

\begin{enumerate}
\item  The first class of vulnerabilities can be used to manipulate the instructions sent to the controller/plant. The adversary can intervene at nodes 1 and 2. At node 1 the adversary modifies the instruction (typically a G-code) sent by the operator. The adversary may intervene at node 2 and tamper with the actuation signal sent to the plant. 
\item The second class of vulnerabilities is the replay attack. At node 4, since the actuation signal is monitored, the replay attack can trick the external observer into thinking that the instructions are executed as per specifications. 
\item  The third class of vulnerabilities arise due to the feedback loops. The internal observer (controller) and the external observer use the machine state and process information to send new instructions.  The adversary may intervene at node 3, 5 and 6 to relay false information on the machine state and process resulting in erroneous feedback control. This sabotages the process of online quality control. 
\item The last class of vulnerabilities is identified at nodes 7 and 8. Node 7 corresponds to the side channel attacks leading to IP theft. Node 8 represents an indirect sabotage of the system in place due to counterfeit production.
\end{enumerate}

In Figure \ref{fig:vnodes_hybrid}, the block $H_{2}(s)$ within the innermost feedback loop is a transfer function block that estimates the machine state (e.g., spindle speed, bed and tool position, laser power) based on the measurements from built-in sensors, such as optical scales and other motion trackers.  The controller is continually tracking the error between the reference signal (generated from the interpretation of the instructions in the G-code) and the feedback signal of the estimated machine state from the hybrid machine tool. The reference signal specifies what the machine state should be at any given point in time as per the instructions in the G-code. The controller sends actuation signals ($\vec{u}$) to the hybrid machine tool that nullifies this error and thus bringing the machine state to the reference state. Injection attacks performed at node 2, include false actuation signals that drive the machine to undesirable states resulting in process faults.  In case of a Man-in-the-Middle attack (replay attack) carried out at node 3, the transfer function block receives incorrect observations (contrary to the actual observations made by the optical scales within the machine) leading to a trail of miscalculations of the estimate of the machine state, error and therefore the actuation signal itself. Therefore, again resulting in the machine being driven to undesirable states and thus eventually faults in the process. 

The block $H_{1}(s)$ in the outer feedback loop estimates the state of the process, based on information from a sensor wrapper \cite{Bhaskar} and generates new instruction sets as required. Typically, the transfer functions tend to be nonlinear operators to fuse information on the nonlinear and nonstationary dynamics underlying the measured signals to detect changes for corrective actions \cite{iquebal2018change} or anticipate anomalies for prognostication and anticipatory control \cite{cheng2015time}. The state of the process is defined in terms of the thermo-mechanical state variables that capture the process that determines transformation of the geometry, morphology, and the microstructure of the part as it is being realized, as well as the health of the machine and its components. Information derived from the sensor wrapper may include thermal history, acoustic emission, and vibrations. The new set of instructions generated based on the estimated process state include reduction of laser power for the DED process if desired melt-pool geometry, thermal history and/or micro structure are not realized, re-manufacturing of layers due to part distortions, and stopping the machine for preventive maintenance due to tool wear. Information on thermal history can be used to predict part deformation during additive manufacturing cycles\cite{Prahalad}. Vibration data in a grinding process can predict surface quality\cite{Bhaskar}. Acoustic emission signals can be used to predict the cutting conditions for orthogonal cutting experiments~\cite{Zimo}. Such applications of the sensory information from the process allow for generation of prognosis-based instructions to the controllers. 

The outer feedback loop tracks the process and serves the purpose of minimizing the process deviation and averting any process anomaly. Attacks on the outer feedback loop have a direct consequence on the inner feedback loop, since instructions generated by the outer feedback loop are direct inputs to the inner feedback loop. Man-in-the-Middle attacks carried out at nodes 4,5 or 6 yield incorrect process state estimations and therefore wrong prognosis leading to generation of incorrect instructions to the controller. Injection attacks at node 1 serve the effect of controllers in the inner feedback loop tracking reference signals generated from the adversary’s instructions, obviating the efforts of the prognosis-based instructions from the external feedback loop.

Side channel attacks at node 7 involve adversaries monitoring the footprint generated by the process. These footprints, for example, can be captured using a microphone that collects the acoustic sounds produced by the machine when in operation \cite{Faruque} or by tapping into the sensor data and other signals in the outer feedback loop. Adversaries that track these footprints from un-monitored channels could reverse engineer the product and create counterfeits which could find their way into the supply chain of critical components. Although the effect of a counterfeited product is not as pronounced in the manufacturing of low volume, high-value customizable parts as is the case where these hybrid machines are put to use, existence of such threats cannot be overlooked. Counterfeit products do not qualify the strict quality standards causing devastation in critical applications. They also sabotage brand reputation. Counterfeiting practices threaten the entire hybrid machine tool that is meticulously put in place with its feedback loops to ensure strict part quality and highlighted as node 8.

\revision{The aforementioned vulnerabilities identified in Figure \ref{fig:vnodes_hybrid} for the hybrid machine tool have been independently exploited in other DM systems such as FDM 3D printers.Various attacks have been devised to exploit the vulnerability and sabotage other such systems. Attacks that resemble the exploitation of the vulnerabilities at the nodes in Figure \ref{fig:vnodes_hybrid} include: \cite{belikovetsky2017dr0wned} demonstrates the modification of G-code (node 1) that resulted in undetectable (node 3) malicious printing sequences being executed; \cite{moore2017implications} develops malicious firmware that modifies original actuation commands to change the 3D printing parameters (node 2,3) that go unnoticed; Attack at nodes 4, 5 and 6, resemble the attacks at node 3, however, they differ in the purpose to which the feedback is put to use. These feedbacks are established for more advanced purposes of sending corrective G-code based on real time process state and sensor data. Attack at these nodes although similar to those at node 3 still remain to be demonstrated. \cite{chhetri2017confidentiality} demonstrates an acoustic emanation based side channel attack  leading to counterfeit by reverse engineering and therefore IP issues (node 7,8). Other similar attacks are presented in the context of the developed taxonomy and discussed as case studies later in section 3.}

\revision{Vulnerabilities outside of the specified schema in Figure \ref{fig:vnodes_hybrid} include those that are innate to any software and data management systems used to interface with the operational technologies, as well as those occuring across a wider supply chain \cite{gupta2020additive} that employs digital manufacturing and the process chains enabled by them. Examples include ransomware outbreaks at TSMC (WannaCry)~\cite{TSMC_OB} in 2018 and Norsk Hyrdo (LockerGoga)~\cite{NH_OB} in 2019 forcing the companies either to halt operations or switch to manual operation costing an estimated \$180M and \$71M respectively.}
\begin{figure*}[t!]
\centering
\includegraphics[width=0.9\textwidth]{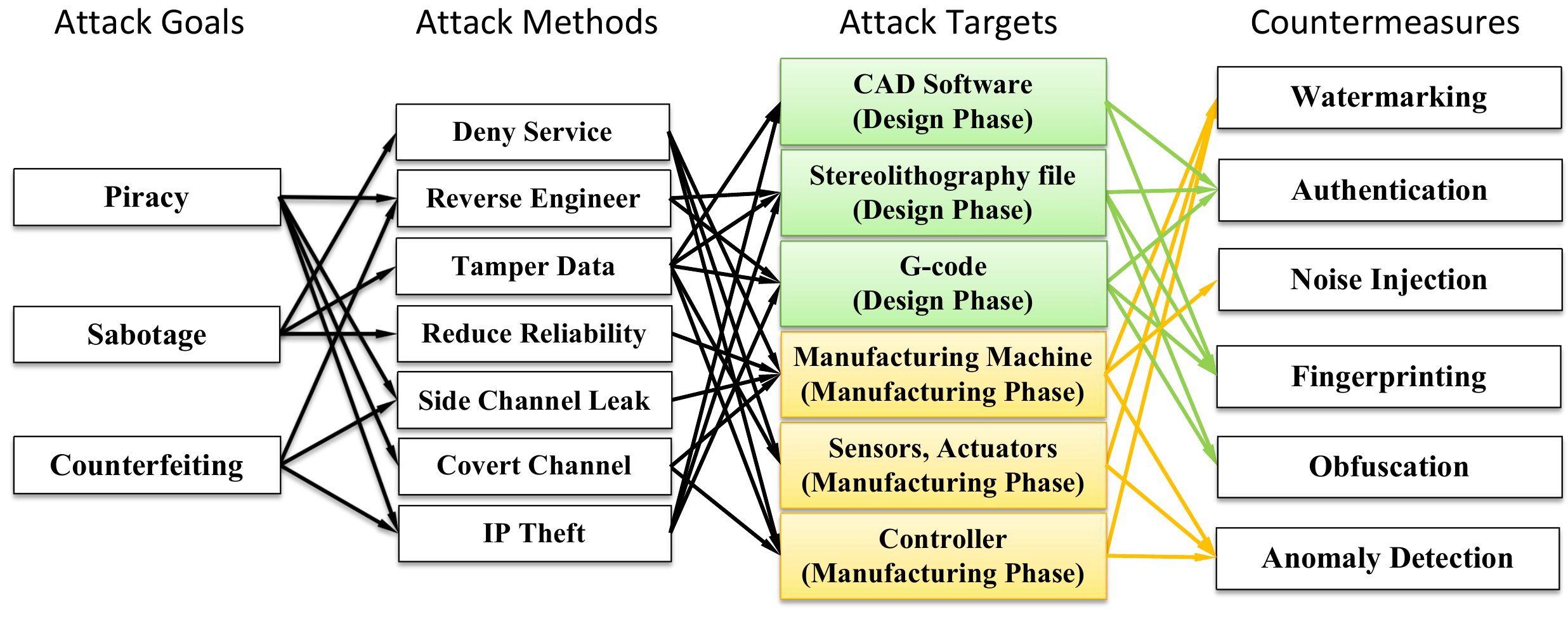}
\caption{\revision{Threat taxonomy and corresponding security measures. The left column (first) shows the goals of attackers, and the second to the left column describes possible attacks, the third column shows the targets of attackers, and the last (right) column shows countermeasures. The arrows from the first column to the second column show how an attacker can achieve different goals using various attacks, and the arrows from the second column to the third column show how each attack can be applied on each target. Lastly, the arrows from the third column to the fourth column show how each component in DM systems can be protected by countermeasures.}}
\label{fig:my_attack-defense}
\end{figure*}

\section{Digital Manufacturing: Taxonomy of Threats}
\label{sec:threats}

Cyber-enablement and interconnectivty of digital supply chain networks introduce threats  including financial theft and theft of IP. Some of the threats are unique to DM including digitally printing dangerous or illegal components, stealing competitor IP (e.g. the design files), modifying them and manufacturing counterfeits or sub-standard components and deny service by taking manufacturing plants or critical parts of the manufacturing plants (e.g. printers) offline. The attackers may have different motivations including (i) nation state actors, (ii) organized criminals, (iii) politically, socially, or ideologically motivated hacktivists, (iv) hackers with financial gain or sabotage intent, (v) competitors, and (vi) malicious insiders. The motivation of the attacker, resources available, and the damage caused in each category can be different and should be a part of the threat analysis.

\subsection{Taxonomy of threats}

Figure~\ref{fig:my_attack-defense} shows a taxonomy of attacks, attack goals, methods, targets and the countermeasures. An attacker can choose their attack method based on their goals and targets. 

\revision{\textbf{Attack Goals:} can be grouped into three classes: }
\begin{enumerate}
    \item \revision{\textbf{Piracy} refers to illegally copying or fabricating a design that violates the copyright of the original design.}
    \item \revision{\textbf{Sabotage} entails introducing defects in the product, damaging machines or interfering with the processes to cause delay or damage.} 
    \item \revision{\textbf{Counterfeiting} attacks are defined as illegal attempts to imitate authentic products.}
\end{enumerate}

\revision{\textbf{Attack Methods} %
can be characterized  into seven categories:}

\begin{enumerate}
    \item \revision{\textbf{Denial of Service} attacks prevent access to the manufacturing systems.}
    \item \revision{\textbf{Reverse Engineering:} Given a file or physical product as the output of on design/manufacturing stage in the supply chain, reconstruct files in a previous step.}
    \item \revision{\textbf{Data Tampering}  refers to tampering of data read/written, stored, sent/received by the manufacturing system.}
    \item \revision{\textbf{Reliability Degradation}: refers to reduction in production yield, on-time performance of systems, and unpredictable decrease in service life of the part.}
    \item \revision{\textbf{Side Channel Leakage} refers to reconstructing the product design and manufacturing conditions  side channel information measured during the manufacturing (e.g., acoustic, thermal, electromagnetic, vibration).}
    \item \revision{\textbf{Covert Channel} attacks are when an insider intentionally sends secret information to the outside receiver while remaining detected or noticed by others.}
    \item \revision{\textbf{IP Theft}: directly stealing digital proprietary information (e.g., design files) from the computers or machines in manufacturing systems. Often, such information can be used for developing competing products.}
\end{enumerate}

\revision{\textbf{Attack Targets}: Based on the supply chain of digital manufacturing system presented in Fig.~\ref{fig:supply_chain}, we first largely classify the targets into design phase targets (marked in green in Fig.~\ref{fig:my_attack-defense}) and manufacturing phase targets (yellow in Fig.~\ref{fig:my_attack-defense}). We identify three targets in each phase as explained below.}
\begin{enumerate}
    \item \revision{\textbf{CAD Software} is widely used to facilitate product design by a single designer or by a collaborative design teams. It can be targeted by an attacker in a data tampering attack, such that the CAD software will not generate the correct files as expected.}
    \item \revision{\textbf{Stereolithography} a.k.a., STL file format is a widely used generic format that describes the surface geometry of a 3-dimensional object by a tessellation scheme. The file resolution can change the product quality.}
    \item \revision{\textbf{G-code} is the numerical control programming language. G-code files define the processing parameters such as tool path, nozzle temperature, laser power, material type, etc. It stores crucial design information and so its integrity and confidentiality are critical.}
    \item \revision{\textbf{Manufacturing Machines} are the physical machines that manufacture the products in the physical world.}
    \item \revision{\textbf{Sensors, Actuators}: In a manufacturing control feedback loop, sensors and actuators are responsible for measuring and driving the physical process, respectively.}
    \item \revision{\textbf{Controllers} in a feedback loop carry out the decision-making process to control the behavior of the machines, and the G-code files define the controller behavior.}
\end{enumerate}

\revision{\textbf{Countermeasures} are in six categories:}

\begin{enumerate}
    \item \revision{\textbf{Watermarking} is a security technique that embeds inseparable and hidden information in signals/files, such that the owner of the original signals/files can use the hidden information to prove its ownership or the authenticity of the signals/files.}
    \item \revision{\textbf{Authentication}  helps identify if they are interacting with the authentic copy of a file/message/identity.}
    \item \revision{\textbf{Noise Injection}: refers to the countermeasures that inject noise signals in its side channel information leakage, so an attacker will not be able to retrieve meaningful secret information from side channel information measurement.}
    \item \revision{\textbf{Fingerprinting} exploits the intrinsic characteristics of designs/machines/processes as a method to uniquely identify designs or products produced by a design file or a manufacturing machine.}
    \item \revision{\textbf{Obfuscation} of design files prevents designs from being understood and thus reverse engineered by malicious attackers. Obfuscation introduces  difficulties for an attacker to reverse engineer a given product.}
    \item \revision{\textbf{Anomaly Detection} can be applied to multiple layers. For example, it can be used on the controller of a manufacturing system to detect whether there are any suspicious sensor readings in the system. It can also be applied to the manufacturing machine itself to detect whether there is anything different from expected behaviors, e.g., by monitoring the side-channel information leakage of the machines. Anomaly detection can also be applied to the network layer to intercept the packages in the network. It can also be applied to the products, and the products will be checked against the specification, especially a few security-critical properties will be checked explicitly.}
\end{enumerate}

The taxonomy presented in  Figure~\ref{fig:my_attack-defense} can be used to develop defenses for various attack scenarios. For example, to prevent an attacker from tampering with the design files (e.g., STL files), a defender can embed identification codes in the design to authenticate the product. If the design is tampered with or reverse engineered, the embedded code will be impacted, and will not match with the correct one.

\revision{}
According to this taxonomy, we classify recent related works in Table~\ref{tab:summary}. We first classify the papers based on whether they focus on attacks or defenses or both. Then the threat models that they consider are identified. In the case that the paper is a survey that covers a variety of threat models, we will leave the threat model field blank. Lastly, we categorize all papers based on the attack methods they presented or based on the defenses. Not surprisingly most papers are focused on defenses. However, in order to develop a defense, the threat model that it targets overwhelmingly indicates that sabotage is the main attack goal and the attacks are launched either to tamper the files or for IP theft. IP theft is a major concern in DM because the design of hardware parts remains the same for many years, even decades. Revision to the designs that have been in place for so long, due to design theft becomes expensive and taxing exercise. A related issue in manufacturing is that a legitimately obtained part can be reverse engineered  and then used for unauthorized or counterfeit production leading to IP theft. The deterrence in such cases lies in the production method that cannot be easily copied or decoded. Although DoS attacks are a major concern in financial and technology sectors, they are not a major concern in the manufacturing sector. This is because in many large manufacturing enterprises, the manufacturing machines are maintained on a separate, protected internal network, which is then securely connected to the internet for software or firmware updates only under supervision when the production activity is not taking place. A growing concern is the manufacturing-unique side channels (e.g., acoustics) and related attacks aided by machine learning to uncover patterns in 
data obtained from the multiple sensing sources such as acoustic, thermal, power meter and security camera sensors.

The threats listed in our taxonomy apply to all manufacturing machines including the hybrid machines. Attackers can sabotage the products by tampering the control signals, or instructions (e.g., the G-Code) from the operators. Attackers can steal design secrets from side channel leaks. To explain the attacks and potential impact of the attacks on various aspects of DM process chain, we present five case studies shown as red rows in Table \ref{tab:summary}.

\begin{table*}[t!]
\centering
\begin{tiny}
\caption{Categorization of DM Security studies. ``DoS'', ``Rev. Engg.'', ``Tamper'', ``Unreliable'', Cov. channel'' stand for ``Denial of Service'', ``Reverse Engineering'', ``Tampering data'', ``Reduce reliability'', and ``covert channel'', respectively. \textcolor{red}{Red rows are attack case studies in section \ref{sec:threats}}. \textcolor{blue}{Blue rows are defense case studies in section \ref{sec:defense}}.}
\label{tab:summary}
\resizebox{0.82\textwidth}{!}{%
\begin{tabular}{|lccccccccccccc|}
\specialrule{1pt}{1pt}{1pt}
 & \multicolumn{2}{c|}{\textbf{}} & \multicolumn{3}{c|}{\textbf{Attack Goals}} & \multicolumn{7}{c|}{\textbf{Attacks}} \\
\textbf{Papers} & \rotatebox{90}{Attacks} & \multicolumn{1}{c|}{\rotatebox{90}{Defenses}} & {\rotatebox{90}{Piracy}} & \rotatebox{90}{Sabotage} & \multicolumn{1}{c|}{\rotatebox{90}{Counterfeit}} & {\rotatebox{90}{DoS}} & {\rotatebox{90}{Rev. Engg.}} & {\rotatebox{90}{Tamper}} & \rotatebox{90}{Unreliable} & \rotatebox{90}{Sidechannel} & \rotatebox{90}{Cov.channel} & \multicolumn{1}{c|}{\rotatebox{90}{IP Theft}}\\ \specialrule{0.8pt}{1pt}{1pt}
\rowcolor[HTML]{DFDFDF} 
Gupta \textit{et al.}~\cite{gupta2020additive} & & \multicolumn{1}{c|}{\checkmark} &  & \checkmark & \multicolumn{1}{c|}{} & & & \checkmark & & & & \multicolumn{1}{c|}{\checkmark}\\
Strurm \textit{et al.}~\cite{sturm2017cyber} & & \multicolumn{1}{c|}{} &  & \checkmark & \multicolumn{1}{c|}{\checkmark} & & & \checkmark & & & & \multicolumn{1}{c|}{}\\
\rowcolor[HTML]{DFDFDF} 
Ranabhat \textit{et al.}~\cite{ranabhat2019optimal} & \checkmark & \multicolumn{1}{c|}{} &  & \checkmark & \multicolumn{1}{c|}{} & & & \checkmark & \checkmark & & & \multicolumn{1}{c|}{}\\
\textcolor{red}{Belikovetsky \textit{et al.}~\cite{belikovetsky2017dr0wned}} & \textcolor{red}{\checkmark} & \multicolumn{1}{c|}{} &  & \textcolor{red}{\checkmark} & \multicolumn{1}{c|}{\textcolor{red}{\checkmark}} & & \textcolor{red}{\checkmark} &  & & & \textcolor{red}{\checkmark} & \multicolumn{1}{c|}{}\\
\rowcolor[HTML]{DFDFDF} 
Yampolskiy \textit{et al.}~\cite{yampolskiy2016using} & & \multicolumn{1}{c|}{\checkmark} &  & \checkmark & \multicolumn{1}{c|}{} & & & & \checkmark & & & \multicolumn{1}{c|}{}\\
Wu \textit{et al.}~\cite{wu2019detecting} & & \multicolumn{1}{c|}{\checkmark} &  & \checkmark & \multicolumn{1}{c|}{} & & & \checkmark & & \checkmark & & \multicolumn{1}{c|}{\checkmark}\\
\rowcolor[HTML]{DFDFDF} 
Chhetri \textit{et al.}~\cite{chhetri2016kcad} & & \multicolumn{1}{c|}{\checkmark} &  & \checkmark & \multicolumn{1}{c|}{} & & & \checkmark & \checkmark & \checkmark & & \multicolumn{1}{c|}{}\\
Desmit \textit{et al.}~\cite{desmit2016cyber} & \checkmark & \multicolumn{1}{c|}{} & \checkmark &  & \multicolumn{1}{c|}{} & & & \checkmark & \checkmark & & & \multicolumn{1}{c|}{\checkmark}\\
\rowcolor[HTML]{DFDFDF} 
Chen \textit{et al.}~\cite{chen2017security} & & \multicolumn{1}{c|}{\checkmark} &  & \checkmark & \multicolumn{1}{c|}{\checkmark} & & & & & & & \multicolumn{1}{c|}{\checkmark}\\
Elhabashya \textit{et al.}~\cite{elhabashya2019cyber} & & \multicolumn{1}{c|}{\checkmark} &  & \checkmark & \multicolumn{1}{c|}{} & & & \checkmark &  & \checkmark & & \multicolumn{1}{c|}{\checkmark}\\
\rowcolor[HTML]{DFDFDF} 
\textcolor{red}{Moore \textit{et al.}~\cite{moore2017implications}} & \textcolor{red}{\checkmark} & \multicolumn{1}{c|}{} &  & \textcolor{red}{\checkmark} & \multicolumn{1}{c|}{} & \textcolor{red}{\checkmark}  & & \textcolor{red}{\checkmark} & \textcolor{red}{\checkmark} & & & \multicolumn{1}{c|}{}\\
Bracho \textit{et al.}~\cite{bracho2018simulation} & & \multicolumn{1}{c|}{\checkmark} &  &  & \multicolumn{1}{c|}{} & & &  & & \checkmark & \checkmark & \multicolumn{1}{c|}{\checkmark}\\
\rowcolor[HTML]{DFDFDF} 
Graves \textit{et al.}~\cite{graves2019characteristic} & \checkmark & \multicolumn{1}{c|}{} & \checkmark & \checkmark & \multicolumn{1}{c|}{} & & & & & & & \multicolumn{1}{c|}{\checkmark}\\
Yampolskiy \textit{et al.}~\cite{yampolskiy2014intellectual} & & \multicolumn{1}{c|}{\checkmark} & \checkmark &  & \multicolumn{1}{c|}{} & & \checkmark & \checkmark & & \checkmark & & \multicolumn{1}{c|}{\checkmark}\\
\rowcolor[HTML]{DFDFDF} 
Chhetri \textit{et al.}~\cite{chhetri2017security} & & \multicolumn{1}{c|}{\checkmark} &  &  & \multicolumn{1}{c|}{} & & \checkmark & \checkmark & & \checkmark & & \multicolumn{1}{c|}{\checkmark}\\
Belikovetsky \textit{et al.}~\cite{belikovetsky2017detecting} & & \multicolumn{1}{c|}{\checkmark} & \checkmark & \checkmark & \multicolumn{1}{c|}{} & & & \checkmark & & \checkmark & & \multicolumn{1}{c|}{}\\
\rowcolor[HTML]{DFDFDF} 
Chhetri \textit{et al.}~\cite{chhetri2017side} & & \multicolumn{1}{c|}{\checkmark} &  &  & \multicolumn{1}{c|}{} & & \checkmark &  & & \checkmark &  & \multicolumn{1}{c|}{\checkmark}\\
Baumann \textit{et al.}~\cite{baumann2017additive} & & \multicolumn{1}{c|}{\checkmark} &  &  & \multicolumn{1}{c|}{} & & \checkmark & \checkmark & & & & \multicolumn{1}{c|}{\checkmark}\\
\rowcolor[HTML]{DFDFDF} 
Wu \textit{et al.}~\cite{wu2018cybersecurity} & \checkmark & \multicolumn{1}{c|}{} &  & \checkmark & \multicolumn{1}{c|}{} & \checkmark & & \checkmark & & & & \multicolumn{1}{c|}{\checkmark}\\
\textcolor{blue}{Gupta \textit{et al.}~\cite{gupta2017obfuscade}} & \textcolor{red}{\checkmark} &  \multicolumn{1}{c|}{\textcolor{blue}{\checkmark}} &  & \textcolor{blue}{\checkmark} & \multicolumn{1}{c|}{\textcolor{blue}{\checkmark}} & & & \textcolor{red}{\checkmark} & & & & \multicolumn{1}{c|}{\textcolor{blue}{\checkmark}}\\
\rowcolor[HTML]{DFDFDF} 
Moore \textit{et al.}~\cite{moore2017power} & \checkmark & \multicolumn{1}{c|}{} & \checkmark & \checkmark & \multicolumn{1}{c|}{} & & &  & & \checkmark &  & \multicolumn{1}{c|}{\checkmark}\\
Tsoutsos \textit{et al.}~\cite{tsoutsos2017secure} &  &  \multicolumn{1}{c|}{\checkmark} &  &  & \multicolumn{1}{c|}{} & &  \checkmark & \checkmark & & & & \multicolumn{1}{c|}{}\\
\rowcolor[HTML]{DFDFDF} 
Belikovetsky \textit{et al.}~\cite{belikovetsky2018digital} & \checkmark & \multicolumn{1}{c|}{} &  & \checkmark & \multicolumn{1}{c|}{} & & \checkmark & \checkmark & \checkmark & \checkmark & & \multicolumn{1}{c|}{\checkmark}\\
Zarreh \textit{et al.}~\cite{zarreh2018cybersecurity} & & \multicolumn{1}{c|}{\checkmark} &  & \checkmark & \multicolumn{1}{c|}{} & \checkmark & & \checkmark & \checkmark & & & \multicolumn{1}{c|}{\checkmark}\\
\rowcolor[HTML]{DFDFDF} 
Miller \textit{et al.}~\cite{miller2018identifying} & & \multicolumn{1}{c|}{\checkmark} &  & & \multicolumn{1}{c|}{} &  & \checkmark & & & \checkmark & \checkmark & \multicolumn{1}{c|}{\checkmark}\\
Chaduvula \textit{et al.}~\cite{chaduvula2018security} &  & \multicolumn{1}{c|}{\checkmark} &  & \checkmark & \multicolumn{1}{c|}{\checkmark} & & & \checkmark & & & & \multicolumn{1}{c|}{}\\
\rowcolor[HTML]{DFDFDF} 
Raban \textit{et al.}~\cite{raban2018foresight} & \checkmark & \multicolumn{1}{c|}{\checkmark} &  & & \multicolumn{1}{c|}{\checkmark} &  & & \checkmark & & & \checkmark & \multicolumn{1}{c|}{\checkmark}\\
\textcolor{blue}{Chen \textit{et al.}~\cite{chen2019embedding}} &  & \multicolumn{1}{c|}{\textcolor{blue}{\checkmark}} &  & & \multicolumn{1}{c|}{\textcolor{blue}{\checkmark}} & & \textcolor{blue}{\checkmark} &  & & & & \multicolumn{1}{c|}{}\\
\rowcolor[HTML]{DFDFDF} 
Yu \textit{et al.}~\cite{yu2020sabotage} & & \multicolumn{1}{c|}{\checkmark} &  & \checkmark & \multicolumn{1}{c|}{} &  & & & & \checkmark & & \multicolumn{1}{c|}{}\\
Hoffman \textit{et al.}~\cite{hoffman2018internet} & & \multicolumn{1}{c|}{\checkmark} &  &  & \multicolumn{1}{c|}{} & & & \checkmark & \checkmark & & & \multicolumn{1}{c|}{\checkmark}\\
\rowcolor[HTML]{DFDFDF} 
Abdulhameed \textit{et al.}~\cite{abdulhameed2019additive} &  & \multicolumn{1}{c|}{\checkmark} &  & \checkmark & \multicolumn{1}{c|}{} &  & & \checkmark & & & & \multicolumn{1}{c|}{}\\
Padmanabhan \textit{et al.}~\cite{padmanabhan2018cybersecurity} & & \multicolumn{1}{c|}{\checkmark} &  & & \multicolumn{1}{c|}{} & & &  & \checkmark & & & \multicolumn{1}{c|}{\checkmark}\\
\rowcolor[HTML]{DFDFDF} 
Prinsloo \textit{et al.}~\cite{prinsloo2019review} &  & \multicolumn{1}{c|}{\checkmark} &  & \checkmark & \multicolumn{1}{c|}{} & \checkmark  & & \checkmark & & \checkmark & \checkmark & \multicolumn{1}{c|}{}\\
Chhetri \textit{et al.}~\cite{chhetri2019tool} & \checkmark & \multicolumn{1}{c|}{\checkmark} &  & & \multicolumn{1}{c|}{} & & &  & &  \checkmark & & \multicolumn{1}{c|}{}\\
\rowcolor[HTML]{DFDFDF} 
Calzado \textit{et al.}~\cite{jimenez2019additive} & & \multicolumn{1}{c|}{\checkmark} &  &  & \multicolumn{1}{c|}{} & & \checkmark & & & & & \multicolumn{1}{c|}{}\\
Yampolskiy \textit{et al.}~\cite{yampolskiy2015security} & \checkmark & \multicolumn{1}{c|}{} &  & \checkmark & \multicolumn{1}{c|}{} & & & \checkmark & & & & \multicolumn{1}{c|}{}\\
\rowcolor[HTML]{DFDFDF} 
Ivanova \textit{et al.}~\cite{ivanova2014unclonable} & & \multicolumn{1}{c|}{} &  &  & \multicolumn{1}{c|}{\checkmark} & & & \checkmark & & & & \multicolumn{1}{c|}{}\\
Bridges \textit{et al.}~\cite{bridges2015cyber} & & \multicolumn{1}{c|}{\checkmark} &  &  & \multicolumn{1}{c|}{} & & \checkmark & \checkmark & & \checkmark & & \multicolumn{1}{c|}{\checkmark}\\
\rowcolor[HTML]{DFDFDF} 
Holland \textit{et al.}~\cite{holland2017copyright} & & \multicolumn{1}{c|}{\checkmark} &  & & \multicolumn{1}{c|}{} & & & & & & & \multicolumn{1}{c|}{\checkmark}\\
Chhetri \textit{et al.}~\cite{chhetri2016poster} & \checkmark & \multicolumn{1}{c|}{} & \checkmark &  & \multicolumn{1}{c|}{} & & & & & \checkmark & & \multicolumn{1}{c|}{\checkmark}\\
\rowcolor[HTML]{DFDFDF} 
Wei \textit{et al.}~\cite{wei2018embedding} & & \multicolumn{1}{c|}{\checkmark} &  &  & \multicolumn{1}{c|}{\checkmark} & & \checkmark &  & & & & \multicolumn{1}{c|}{}\\
Wu \textit{et al.}~\cite{wu2017detecting} & & \multicolumn{1}{c|}{\checkmark} &  & \checkmark & \multicolumn{1}{c|}{} & & & \checkmark & & & & \multicolumn{1}{c|}{}\\
\rowcolor[HTML]{DFDFDF} 
Vincent \textit{et al.}~\cite{vincent2015trojan} & & \multicolumn{1}{c|}{\checkmark} &  & \checkmark & \multicolumn{1}{c|}{} & \checkmark & & & \checkmark & \checkmark & & \multicolumn{1}{c|}{\checkmark}\\
Riel \textit{et al.}~\cite{riel2017integrated} & & \multicolumn{1}{c|}{\checkmark} &  &  & \multicolumn{1}{c|}{} & \checkmark & &  & & \checkmark & \checkmark & \multicolumn{1}{c|}{}\\
\rowcolor[HTML]{DFDFDF} 
Ren \textit{et al.}~\cite{ren2017cyber} & & \multicolumn{1}{c|}{\checkmark} &  &  & \multicolumn{1}{c|}{} & \checkmark & & \checkmark & & & \checkmark & \multicolumn{1}{c|}{}\\
He \textit{et al.}~\cite{he2016security} & & \multicolumn{1}{c|}{\checkmark} &  & \checkmark & \multicolumn{1}{c|}{} & \checkmark & &  & & & \checkmark & \multicolumn{1}{c|}{\checkmark}\\
\rowcolor[HTML]{DFDFDF} 
Wu \textit{et al.}~\cite{wu2018establishment} & \checkmark & \multicolumn{1}{c|}{\checkmark} &  & \checkmark & \multicolumn{1}{c|}{} & & \checkmark & \checkmark & \checkmark & & & \multicolumn{1}{c|}{}\\
Fey \textit{et al.}~\cite{fey20173d} & & \multicolumn{1}{c|}{\checkmark} &  & \checkmark & \multicolumn{1}{c|}{} & & & & \checkmark & & & \multicolumn{1}{c|}{\checkmark}\\
\rowcolor[HTML]{DFDFDF} 
Elhabashy \textit{et al.}~\cite{elhabashy2019cyber} & & \multicolumn{1}{c|}{\checkmark} &  & \checkmark & \multicolumn{1}{c|}{} & & & \checkmark & & & & \multicolumn{1}{c|}{}\\
Slaughter \textit{et al.}~\cite{slaughter2017ensure} & \checkmark & \multicolumn{1}{c|}{\checkmark} &  & \checkmark & \multicolumn{1}{c|}{} & & & \checkmark & & \checkmark & & \multicolumn{1}{c|}{}\\
\rowcolor[HTML]{DFDFDF} 
Satchidanandan \textit{et al.}~\cite{satchidanandan2016secure} & & \multicolumn{1}{c|}{\checkmark} &  & \checkmark & \multicolumn{1}{c|}{} & & & \checkmark & & & & \multicolumn{1}{c|}{}\\
Satchidanandan \textit{et al.}~\cite{satchidanandan2018control} &  & \multicolumn{1}{c|}{\checkmark} &  & \checkmark & \multicolumn{1}{c|}{} & & & \checkmark & & & & \multicolumn{1}{c|}{}\\
\rowcolor[HTML]{DFDFDF} 
\textcolor{red}{Woollaston~\cite{honda}} & \textcolor{red}{\checkmark} & \multicolumn{1}{c|}{} &  & \textcolor{red}{\checkmark} & \multicolumn{1}{c|}{} & \textcolor{red}{\checkmark} & & & & & & \multicolumn{1}{c|}{}\\
\textcolor{blue}{Satchidanandan \textit{et al.}~\cite{proceedings}} &  & \multicolumn{1}{c|}{\textcolor{blue}{\checkmark}} &  & \textcolor{blue}{\checkmark} & \multicolumn{1}{c|}{} & & & \textcolor{blue}{\checkmark}  & & & & \multicolumn{1}{c|}{}\\
\rowcolor[HTML]{DFDFDF} 
\textcolor{blue}{Behera \textit{et al.}~\cite{behera2019system}} &  & \multicolumn{1}{c|}{\textcolor{blue}{\checkmark}} &  & \textcolor{blue}{\checkmark} & \multicolumn{1}{c|}{} & &  & \textcolor{blue}{\checkmark} & & & & \multicolumn{1}{c|}{}\\
Wu \textit{et al.}~\cite{wu2020alert} &  & \multicolumn{1}{c|}{\checkmark} &  & \checkmark & \multicolumn{1}{c|}{} & &  & \checkmark & & & & \multicolumn{1}{c|}{}\\
\rowcolor[HTML]{DFDFDF} 
Yanamandra \textit{et al.}~\cite{yana2020} & \checkmark & \multicolumn{1}{c|}{} & \checkmark & & \multicolumn{1}{c|}{\checkmark} & &  & & & & & \multicolumn{1}{c|}{\checkmark}\\
Do \textit{et al.}~\cite{do2016data} & \checkmark & \multicolumn{1}{c|}{} & \checkmark & \checkmark & \multicolumn{1}{c|}{} & &  & \checkmark & & & & \multicolumn{1}{c|}{\checkmark}\\
\rowcolor[HTML]{DFDFDF} 
Gao \textit{et al.}~\cite{gao2018watching} &  & \multicolumn{1}{c|}{\checkmark} & \checkmark & & \multicolumn{1}{c|}{\checkmark} & &  & & & & & \multicolumn{1}{c|}{\checkmark}\\
Chhetri \textit{et al.}~\cite{chhetri2018information} &  & \multicolumn{1}{c|}{\checkmark} & \checkmark & & \multicolumn{1}{c|}{} & &  & & & \checkmark & & \multicolumn{1}{c|}{}\\
\rowcolor[HTML]{DFDFDF} 
Chhetri \textit{et al.}~\cite{chhetri2017confidentiality} & \checkmark & \multicolumn{1}{c|}{} & \checkmark & & \multicolumn{1}{c|}{} & &  & & & \checkmark & & \multicolumn{1}{c|}{}\\
Chen \textit{et al.}~\cite{chen2019obfuscation} &  & \multicolumn{1}{c|}{\checkmark} &  & & \multicolumn{1}{c|}{\checkmark} & & \checkmark & \checkmark & &  & & \multicolumn{1}{c|}{}\\
\rowcolor[HTML]{DFDFDF} 
Song \textit{et al.}~\cite{song2018my} &  & \multicolumn{1}{c|}{\checkmark} & & & \multicolumn{1}{c|}{\checkmark} & & \checkmark & & & & & \multicolumn{1}{c|}{\checkmark}\\
Song \textit{et al.}~\cite{song2016my} & \checkmark & \multicolumn{1}{c|}{} & \checkmark &  & \multicolumn{1}{c|}{\checkmark} & & & & & \checkmark & & \multicolumn{1}{c|}{}\\
\rowcolor[HTML]{DFDFDF} 
Al Faruque \textit{et al.}~\cite{al2016forensics} & \checkmark & \multicolumn{1}{c|}{} &\checkmark & & \multicolumn{1}{c|}{\checkmark} & &  & & & \checkmark & & \multicolumn{1}{c|}{}\\
\specialrule{0.8pt}{1pt}{1pt}
\end{tabular}
}
\end{tiny}
\end{table*}

\subsection{Case Study 1 –Dr0wned attack on AM~\cite{belikovetsky2017dr0wned}}

\begin{figure*}[htb]
    \centering
    \subfloat[\label{fig:blade1}]{%
       \includegraphics[width=0.25\textwidth]{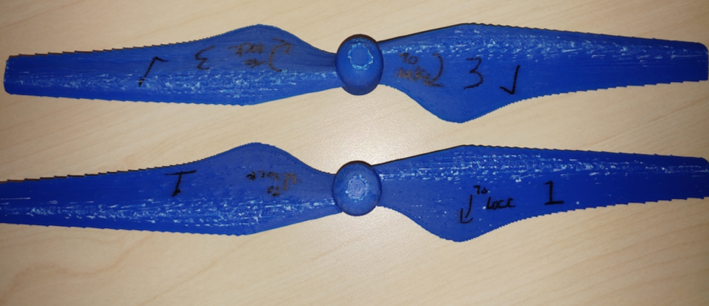}
     }
     \subfloat[\label{fig:blade2}]{%
       \includegraphics[width=0.25\textwidth]{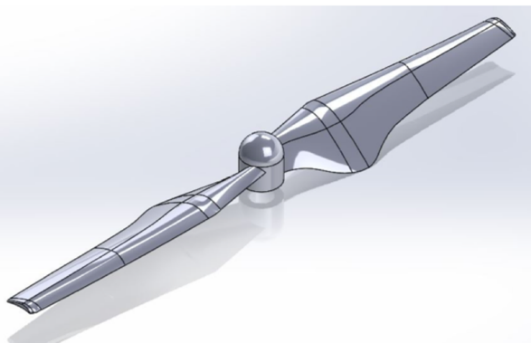}
     }
     \subfloat[\label{fig:blade3}]{%
       \includegraphics[width=0.5\textwidth]{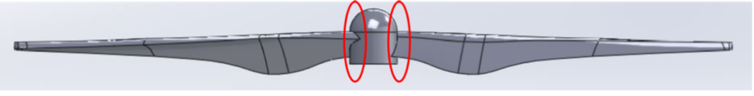}
     }
    \caption{(a) Two 3D printed propellers. One of is defective. (b) CAD model of the design. (c) Design is compromised at the joints causing in-service failure.~\cite{belikovetsky2017dr0wned}}
    \label{fig:my_label}
\end{figure*}

Informed by taxonomy of  Figure~\ref{fig:my_attack-defense}, the goal of this attack was sabotage. The attack was conducted to reduce reliability of the part, and the attack target was design files. This attack on a 3D printer deliberately introduced defects into the part during printing~\cite{belikovetsky2017dr0wned}. The controller PC connected to the 3D printer was compromised by exploiting an un-patched vulnerability in WinRAR. The attack decreased the fatigue life of a quadcopter propeller causing a mid-flight failure by manipulating the part geometry (an example shown in Figure~\ref{fig:blade2}). The attack was  executed in three stages: The attacker compromises the Controller PC, developed a counterfeit design similar to the original design, and replaced the original design file on the victim’s PC with the counterfeit design file with the manipulations shown in Figure~\ref{fig:blade3}. A reverse shell backdoor was installed on the PC, which was used to submit jobs to the 3D printer. This allowed the malicious software to take over the 3-D printer and execute commands by the hacker. 
According to our taxonomy, a variety of defenses can be applied to this scenario. Although the attacker exploited a software vulnerability, the sabotage was detected  by rigorously testing the part. 

\subsection{Case Study 2: Cyberattack on Honda auto  plant~\cite{honda}}

\begin{figure*}[ht]
    \centering
    \includegraphics[width=\textwidth]{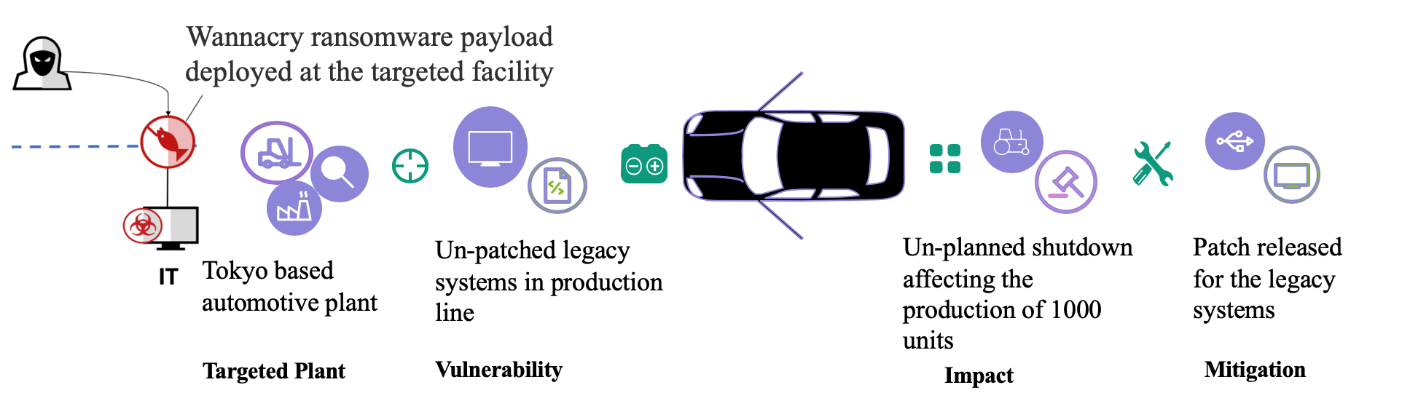}
    \caption{WannaCry cyberattack on the Honda automotive plant computer network~\cite{honda}}.   
    \label{fig:honda}
\end{figure*}

Honda' Tokyo-based automotive production plant was forced to go offline by the self-propagating malware WannaCry impacting the production of about 1000 vehicles~\cite{honda}. The WannaCry malware infected hundreds of thousands of computers worldwide by exploiting vulnerabilities in un-patched legacy systems \cite{wannacry}. The plant was shut down for 48 hours to recover operations and data, as both the ICS and IT networks were impacted~\cite{honda}. As shown in Figure~\ref{fig:honda} the  ransomware got deployed in the plant computer network using a backdoor in an older un-patched version of the windows OS and then infected all systems in the network. %
According to our taxonomy in Figure~\ref{fig:my_attack-defense}, the attacker in this case launched a DoS attack on the automotive plant by infecting and tampering their controller computers in the control network.

\subsection{Case Study 3: Additive Manufacturing Firmware Attack~\cite{moore2017implications}}

Attackers may target the firmware of 3D printer. If the firmware is compromised, attackers can sabotage the system by either modifying the control or deny the service of the machines. The attacker's strategy is to exploit the firmware in order to selectively affect the integrity of printed artifacts; this approach is particularly effective in case random sample testing is applied after the artifact is printed, as it increases the chance of bypassing detection. Furthermore, any intervention to the printer firmware (especially at the bootloader level) can make the attack persistent.

There are different tactics an attacker can employ to infect the printer firmware. Most 3D printers and hybrid manufacturing platforms support Internet connectivity to allow remote management or troubleshooting from the manufacturer, as part of a service-level agreement with the end-users. In this case, attackers can exploit vulnerabilities in the network services running on the printer and eventually escalate their privileges on the printer. This privilege escalation can be exploited to update the printer with infected firmware, in case signed firmware updates are not supported. 
Another attack vector that may be exploited, is the input file parser within the printer. In cases where the firmware processes tool path input files (e.g. G-code files), any input sanity vulnerability may allow memory corruption and execution flow hijacking. In this case, attackers can inject malicious routines through input files, or reuse existing code within the firmware memory space. 

As soon as an attacker has infected the printer firmware, they can easily control the actuators of the printer (e.g., print head motors, extruder valves or laser operation). By controlling these actuators in a judicious fashion, attackers can inject physical property attacks~\cite{moore2017implications}. Furthermore, attackers can also perform a Denial of Service (DoS) attack to the printer so that legitimate users can no longer use the 3D print service.

\subsection{Case Study 4: Dissolvable support material \cite{gupta2017obfuscade}}\label{ss:dissolve}
This attack is applicable to multihead/multimaterial printers, where support material can be printed in addition to the build material. Typically, the support material is dissolvable and as soon as the part is printed, it is submerged into an oxidizer (e.g., acid) to separate it from the build material. The attack consists of maliciously replacing build material in the interior details of the 3D part with support material. Then, as soon as the print is complete and the solvent removes all support material, it would also carve hollow spaces within the part, where original build material was replaced. The effect of this attack is to reduce the structural integrity of the part, since the internal structure will no longer be solid. According to our taxonomy in Figure~\ref{fig:my_attack-defense}, this attack can be classified either as sabotage on DM machine or on the design files set up for multimaterial printing in order to reduce the reliability of the products.

\begin{table*}
\centering
\caption{\revision{Survey and Taxonomy of taxonomies. The green, yellow, and gray columns represent computer security, electronic manufacturing system security, and mechanical manufacturing system security, respectively.}}
\resizebox{\textwidth}{!}{
\label{tab:taxonomy}
\begin{tabular}{|c|l|>{\columncolor{lightgreen}}c|>{\columncolor{yellow}}c|>{\columncolor{yellow}}c|>{\columncolor{yellow}}c|>{\columncolor{yellow}}c|>{\columncolor{lightgray}}c|>{\columncolor{yellow}}c|>{\columncolor{lightgray}}c|>{\columncolor{lightgray}}c|>{\columncolor{lightgray}}c|>{\columncolor{lightgray}}c|>{\columncolor{lightgray}}c|>{\columncolor{yellow}}c|>{\columncolor{lightgray}}c|}
\hline

\multicolumn{2}{|c|}{Papers $\longrightarrow$}& \cite{Landwehr1994taxonomy} & \cite{karri2010trustworthy} & \cite{rostami2014primer} & \cite{bhunia2014hardware} & \cite{ghosh2014secure} & \cite{yampolskiy2016using} & \cite{xiao2016hardware} & \cite{gupta2017obfuscade} & \cite{pan2017taxonomies} & \cite{wu2017taxonomy} & \cite{yampolskiy2018security} & \cite{elhabashy2019cyber} & \cite{hoque2020automated} & Ours   \\ \cline{1-2}

\multicolumn{2}{|c|}{Timeline $\longrightarrow$} & 1994 & 2010 & 2014 & 2014 & 2014 & 2016 & 2016 & 2017 & 2017 & 2017 & 2018 & 2019 & 2020 & 2020\\ \hline

\parbox[t]{2mm}{\multirow{6}{*}{\rotatebox[origin=c]{90}{Attacks} }} & Sabotage (Product) & \checkmark        &    \checkmark     & \checkmark & \checkmark & \checkmark & \checkmark &  & \checkmark & \checkmark & \checkmark & \checkmark & \checkmark & \checkmark & \checkmark\\ \cline{2-16}

& Sabotage (Machine) &         &  & & & & \checkmark & & & \checkmark & \checkmark & \checkmark & \checkmark & & \checkmark \\ \cline{2-16} 
                          
& Sabotage (Environment) &         &   & &     &  & \checkmark  &  & & & \checkmark & \checkmark & & &\\ \cline{2-16}
                          
& Information Leakage &    \checkmark     &  \checkmark       & \checkmark  & \checkmark &  & & & \checkmark & \checkmark & \checkmark & \checkmark & &\checkmark & \checkmark\\ \cline{2-16} 
                         
& Piracy &         &     & \checkmark  &  & \checkmark & & & \checkmark & \checkmark & \checkmark & \checkmark & & & \checkmark\\ \cline{2-16}
                         
& Counterfeit &         &       & \checkmark  &  & \checkmark & & & \checkmark & & &\checkmark & & & \checkmark \\\hline
                         
\parbox[t]{2mm}{\multirow{9}{*}{\rotatebox[origin=c]{90}{Countermeasures}}} &     Obfuscation         &         &         & \checkmark & \checkmark & & & \checkmark & & & & & &  & \checkmark \\ \cline{2-16} 

 &    Watermarking         & &        &    \checkmark     &  &  &  & & & & & & & & \checkmark \\ \cline{2-16}
                          
&    Authentication                 &         &  & \checkmark & & & & & & & & & & & \checkmark \\ \cline{2-16}
                        
&    Noise Injection        &         &      & \checkmark    &  & & & & & & & & & & \checkmark  \\ \cline{2-16}

& Post-Deployment Monitoring & & & & \checkmark & & & \checkmark & & & & & & &\\ \cline{2-16}

& Anomaly Detection & & & & \checkmark & & & \checkmark & & & & & & & \checkmark \\ \cline{2-16}

& Split Manufacturing & & & \checkmark &  & & & \checkmark & & & & & & &\\ \cline{2-16}

&     Fingerprinting         &         &         & \checkmark &  & & & & & & & & & & \checkmark \\ \hline
                          
\parbox[t]{2mm}{\multirow{5}{*}{\rotatebox[origin=c]{90}{Metrics}}}  &   Attempts to find secret     &         & &    \checkmark    &  &  &  & & & & & & & &  \\ \cline{2-16}
            
&     \# of Collisions     &         &         & \checkmark & & & & & & & & & &  &  \\ \cline{2-16}
            
&     Amount of Info. Leakage     &         &         & \checkmark &  &  & & & & & & & & & \\ \cline{2-16}

&     Detection Probability     &         &         & \checkmark &  & & & & & & & & &  &  \\ \cline{2-16}

&     False Positive Rate     &         &         & \checkmark &  &  & & & & & & & & & \\ \hline
\end{tabular}}
\end{table*}

\revision{
\section{Survey and Taxonomy of Taxonomies in Digital (Manufacturing) Systems}\label{sec:comparison}}

\revision{Many relevant cybersecurity taxonomies have been proposed in the past, e.g., in the area of general cybersecurity~\cite{Landwehr1994taxonomy}, electronic manufacturing (supply chain) security~\cite{karri2010trustworthy,rostami2014primer,bhunia2014hardware,ghosh2014secure,xiao2016hardware,hoque2020automated}, and mechanical manufacturing system security~\cite{yampolskiy2016using,gupta2017obfuscade,wu2017taxonomy,yampolskiy2018security,elhabashy2019cyber}. In this section, we will go through the history and present a comprehensive study of security taxonomies for manufacturing systems. A comparison is shown in Table~\ref{tab:taxonomy}.}

\revision{A taxonomy of malicious computer software was introduced in~\cite{Landwehr1994taxonomy}. In the early days of cybersecurity research, the main goals of cyber attacks were to either take over the control of a computer or steal secret information from a computer system. They are still the main focuses of security research nowadays. However, with the introduction of cyber-physical systems, the scope of attacks has been significantly extended.}

\revision{In 2010, the threat landscape  extended to the underlying hardware of a computer system, and Karri \textit{et al.} proposed a taxonomy of hardware Trojans in ICs~\cite{karri2010trustworthy}. The taxonomy shows how a chip can be maliciously designed or fabricated to jeopardize the security of the whole computer system.}

\revision{Rostami \textit{et al.} presented a taxonomy covering a much broader scope of hardware supply chain security~\cite{rostami2014primer}. The taxonomy includes a variety of attacks, including sabotaging the integrated circuits (IC) and computer systems, stealing information, IC design piracy, and IC counterfeiting. In addition to attacks, it also discusses countermeasures and security metrics. Most importantly, in their taxonomy, the connections between countermeasures and corresponding attacks are presented clearly. This provides a comprehensive overview of the field, which greatly facilitates the readers in understanding how to defend against certain attacks. Our taxonomy follows the structure presented in~\cite{rostami2014primer}, as a comprehensive overview of the field of cybersecurity of DM systems.}

\revision{In 2014, Bhunia \textit{et al.} extended the taxonomy of hardware Trojans in electronics manufacturing and added a classification scheme for countermeasures of hardware Trojans~\cite{bhunia2014hardware}. The general categories of countermeasures include runtime monitoring, anomaly detection, and design for trust techniques.}

\revision{Also, in 2014, Ghosh \textit{et al.} extended the scope of IC manufacturing security to printed circuit board manufacturing security~\cite{ghosh2014secure}. They proposed an attack taxonomy, which includes malicious modification during manufacturing, piracy, and product counterfeiting issues.}

\revision{In 2016, Yampolskiy \textit{et al.} analyzed the possibility of turning an additive manufacturing system to a weapon, which can cause physical damages, injuries or death, and environmental contamination~\cite{yampolskiy2016using}. In this analysis, a taxonomy was proposed to analyze the kind of elements that can be compromised in the system, and how an attacker can manipulate other elements in the system through the compromised element. One aspect not often discussed in other related surveys is maliciously tampered source materials that can introduce potential hazards or risks to the system. Also, since the focus of the paper was to study the feasibility of weaponizing additive manufacturing systems, secret information leakage was not covered by the taxonomy at all~\cite{yampolskiy2016using}.}

\revision{In~\cite{xiao2016hardware}, Xiao \textit{et al.} compiled a decade of research on the topic of hardware Trojans. They proposed a comprehensive taxonomy of countermeasures of hardware Trojans to categorize countermeasures. The three main categories of hardware Trojan countermeasures are anomaly detection, split manufacturing, and design for trust.}

\revision{In \cite{gupta2017obfuscade}, Gupta \textit{et al.} presented a taxonomy summarizing the potential attacks and risks of additive manufacturing systems. In their taxonomy, they classified attacks on additive manufacturing based on the step (when), means (how), outcome (what), intent (why), and abstraction (where) of the attacks.}

\revision{Pan \textit{et al.} presented two taxonomies in~\cite{pan2017taxonomies}: one is the threat taxonomy for manufacturing systems, and the other is for quality control systems. Interestingly, the threat taxonomy for manufacturing systems is constructed as a chain for attack development, starting from possible vulnerabilities, then vulnerabilities can be exploited by attack vectors to achieve attack goals on the target. Also, the goals are defined as abstract security properties, including confidentiality, integrity, and availability. Thus, it may not be well connected with readers who do not have cybersecurity background.}

\revision{Wu \textit{et al.} introduced a  taxonomy of cross domain attacks on cyber manufacturing systems~\cite{wu2017taxonomy}. Similar to other taxonomies, the taxonomy in~\cite{wu2017taxonomy} consists of four dimensions: attack vectors, attack impacts, attack methods, and attack consequences. Remarkably, the authors highlighted the domains of different attacks, either in cyber or physical domain. }

\revision{Yampolskiy \textit{et al.} proposed a detailed taxonomy for the security threats in additive manufacturing systems~\cite{yampolskiy2018security}. It first classified all the security threats based on the attackers' goals into two categories: theft of technical data and sabotage. Then the attack targets and attack methods for these two attack goals are presented in two taxonomies separately. The taxonomies classified the attack targets and methods in very fine-grained details, and the descriptions are specific to additive manufacturing. This significantly helps readers understand the whole taxonomy, but it also limits its applicability to other manufacturing systems. The proposed taxonomy is at a more general level than that in~\cite{yampolskiy2018security}; we hope that this taxonomy applies to a wider range of manufacturing systems. } 

\revision{Elhabashy \textit{et al.} proposed an attack taxonomy of production systems~\cite{elhabashy2019cyber}. Their taxonomy and ours have the same structure, i.e., we all classify the security threats on manufacturing/production systems based on attack goals/objectives, attack methods, attack targets/locations. Since Elhabashy \textit{et al.} analyzed the systems from a quality control perspective, they only considered security threats, which will lead to low-quality/ altered products. Comparing with the one in~\cite{elhabashy2019cyber}, our taxonomy in Fig.~\ref{fig:my_attack-defense} has broader coverage in terms of the attack goals/objectives, i.e., we include security threats (Counterfeiting and Piracy) that can potentially steal sensitive information from manufacturing systems. Consequently, more attack methods are included in our taxonomy, e.g., reverse engineering and side-channel leaks.}
\revision{In~\cite{hoque2020automated}, a detailed taxonomy of Trojan attacks on printed circuit board (PCB) was presented. The primary purpose of  Trojans in PCBs is either function tampering or information leakage from the PCBs.}

\revision{\textbf{Our Taxonomy} is developed based on a seminal work that introduced a taxonomy of hardware security threats~\cite{rostami2014primer}. Similar to other related attack taxonomy on (additive) manufacturing systems mentioned above, we also identify attack goals, methods, and targets as important dimensions to categorize and understand attacks on digital manufacturing systems. In addition, we introduce countermeasures in the taxonomy following the approach used in~\cite{rostami2014primer}, so that one can use our taxonomy to quickly identify possible countermeasures for an attack of concern to him/her. We highlight the connections between adjacent dimensions to help readers build a knowledge graph of cybersecurity of digital manufacturing systems. From table~\ref{tab:taxonomy}, we also notice that our taxonomy does not include post-deployment monitoring and split manufacturing as countermeasures, because, to the best of our knowledge, there are no existing works that take these two approaches to protect digital manufacturing systems. However, these may also present new directions for developing novel countermeasures for digital manufacturing systems.}

\section{Digital Manufacturing: Cyberphysical Countermeasures}
\label{sec:defense}
This section presents five case studies \revision{(marked in blue in Table~\ref{tab:summary})} of manufacturing-unique defenses spanning watermarking of controllers used in a range manufacturing settings, design obfuscation, part identification and provenance checking using embedded codes, authentication of designs in the signal processing domain, and an epidemiological approach to manufacturing IoT device security by leveraging their  diversity.

\subsection{Securing Manufacturing Controllers via Dynamic Watermarks\cite{satchidanandan2018control,satchidanandan2016secure}}\label{introduction}

\revision{As outlined in the foregoing, the sensors, actuators and control laws
play a critical role  in DM. systems pertinent to both discrete manufacturing and continuous process  industry. Discrete manufacturing is concerned with manufacture or assembly of discrete units. 
In process industries, the production processes are continuous and batches are indistinguishable \cite{IISE:UnknownDate}. In either case, the production process often depends on maintaining the compositions, temperatures, feed rates, pressures, the levels of tanks, or flow rates, etc. 
The regulation of all the required variables is done through a feedback control loop that senses the relevant output variables and calculates what actuation commands to apply.}

\begin{figure}[t!]
  \centering
  	\centering
    \includegraphics[width=0.45\textwidth]{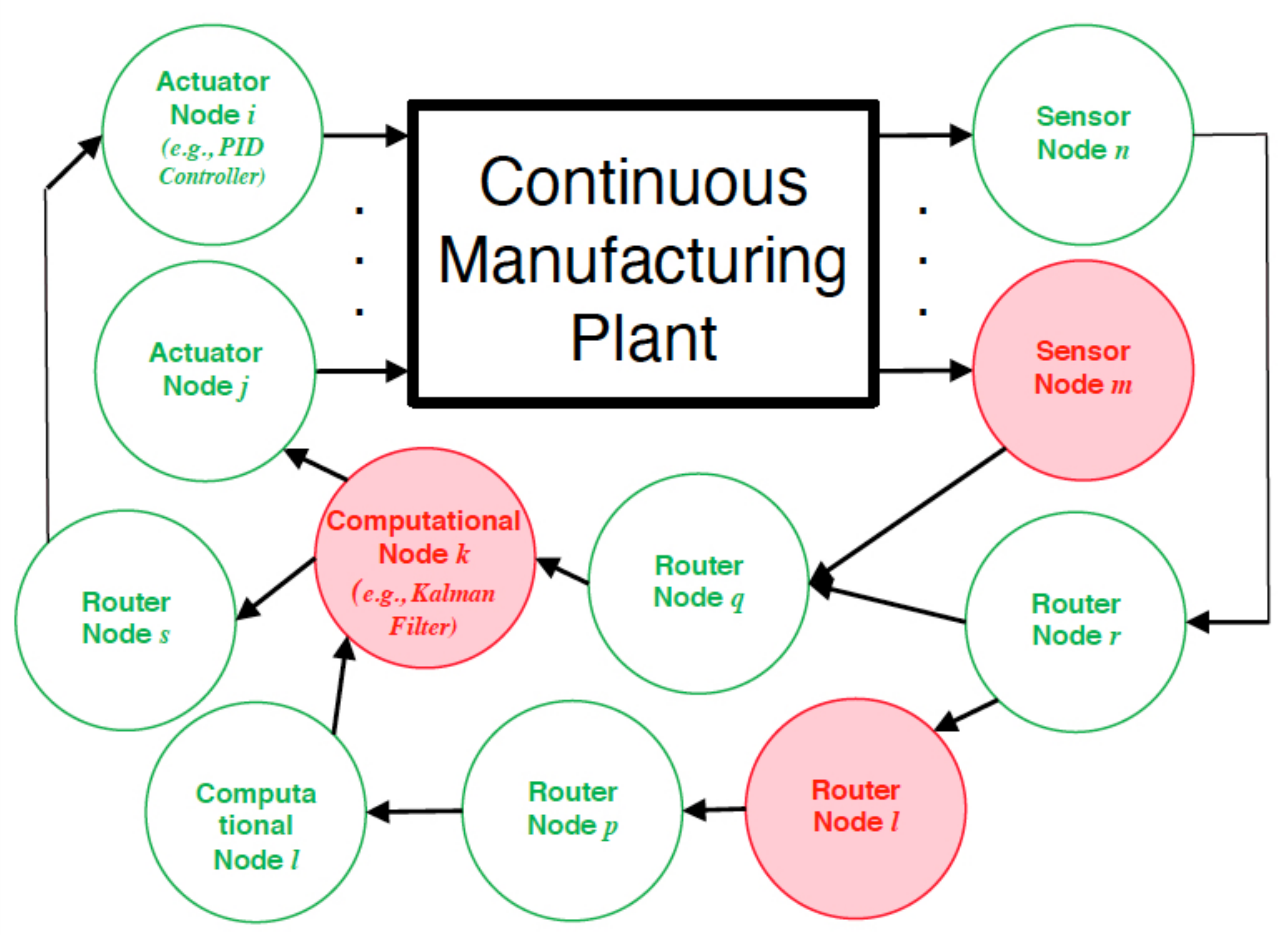}
\caption{\small A manufacturing plant with some subverted nodes.}
\label{fig:overall-problem}
\vspace{-0.1cm}
\end{figure}

The measurements made by the sensors typically travel over a communication network.
The measurements may also be processed at nodes in the network either for fusing information or
for performing computations to support the control law. The problem of cybersecurity arises since sensor measurements or other information traveling over the communication network may be intercepted en route and altered. It is also possible that in distributed control systems, the sensors may be compromised to report false measurements.
Therefore, for securing the manufacturing processes, it is critical to address the security
of the overall distributed control system.
Figure \ref{fig:overall-problem} depicts a manufacturing plant with some 
compromised nodes in the
feedback loops.

\begin{figure}[b!]
  \centering
  	\centering
\includegraphics[width=\columnwidth]{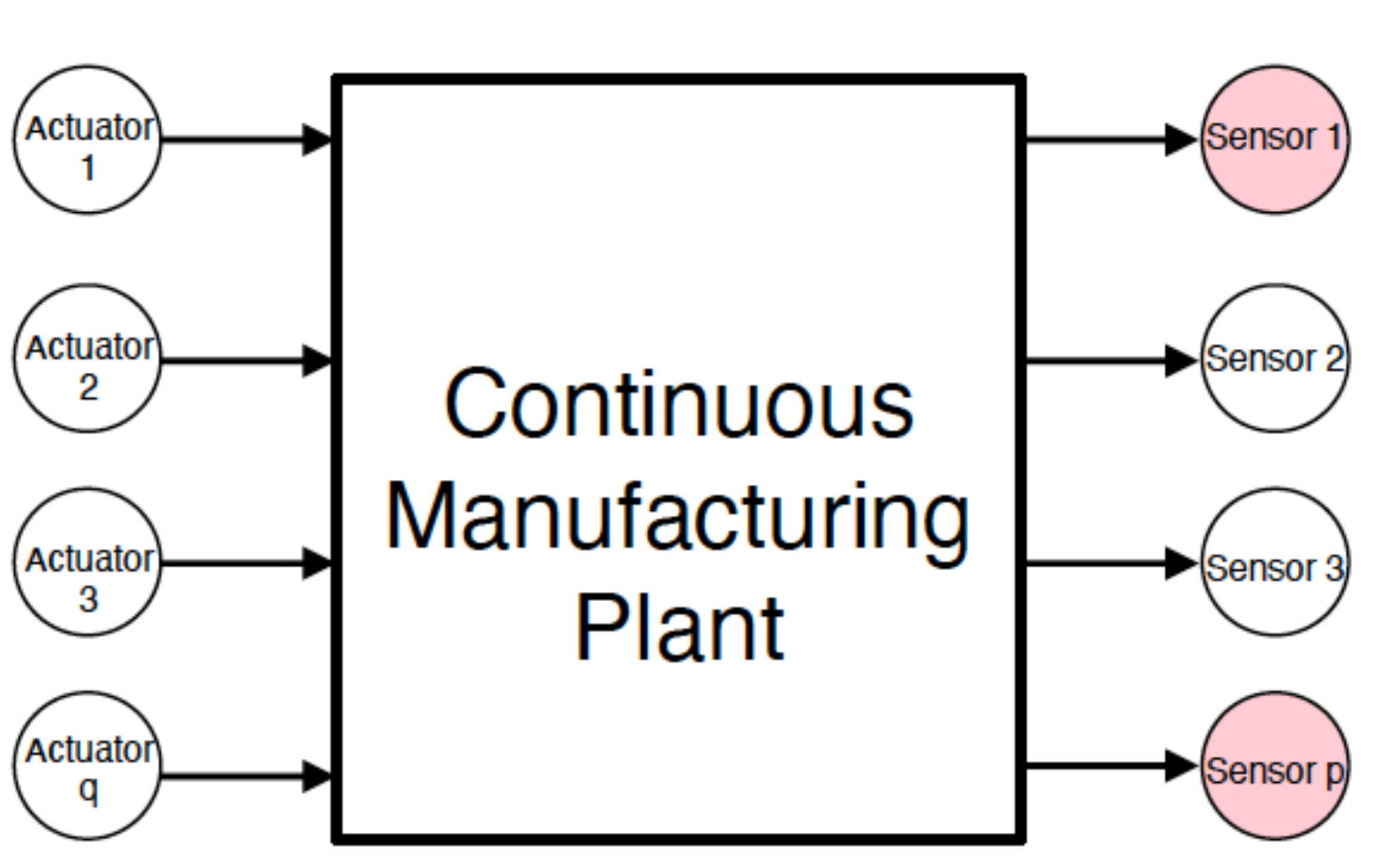}
\caption{\small The abstraction of a manufacturing plant with compromised sensors.}
\label{fig:AbstractionsOfSensorsAndActuatorsCompromised}
\vspace{-0.1cm}
\end{figure}

One can unify all the cases via a simple abstraction where just sensors are compromised, as indicated in Figure \ref{fig:AbstractionsOfSensorsAndActuatorsCompromised}.
Wherever the corruption of the measurements may have taken place, one can just suppose that the sensor has been compromised.

\begin{figure}[ht] 
  \centering
  	\centering
    \includegraphics[width=\columnwidth]{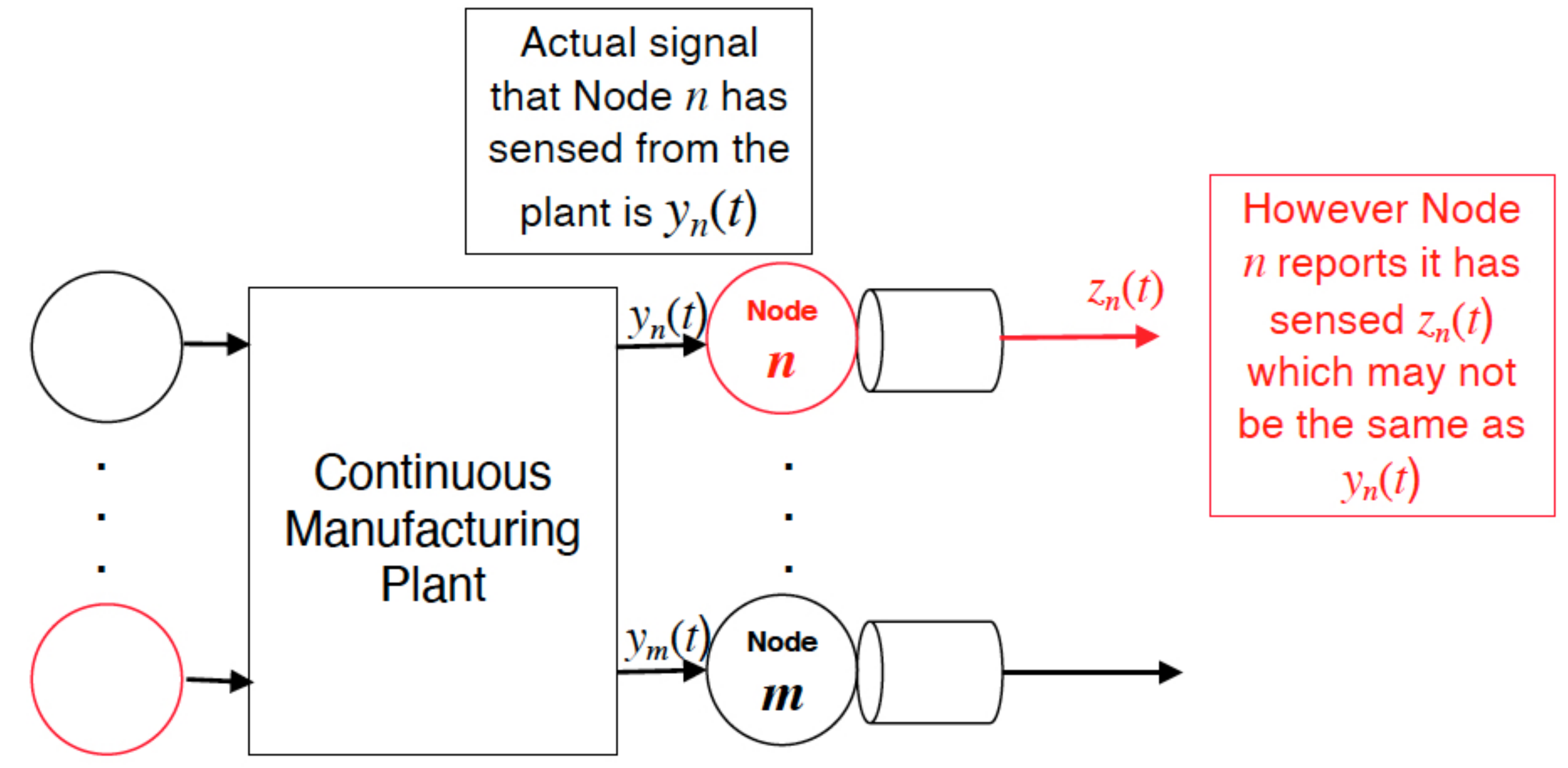}
  \quad
\caption{The malicious behavior of sensor nodes.
}
\label{fig:Malicious-Sensor}
\vspace{-0.2cm}
\end{figure}

\begin{figure}[b!]
  \centering
  	\centering
\includegraphics[width=\columnwidth]{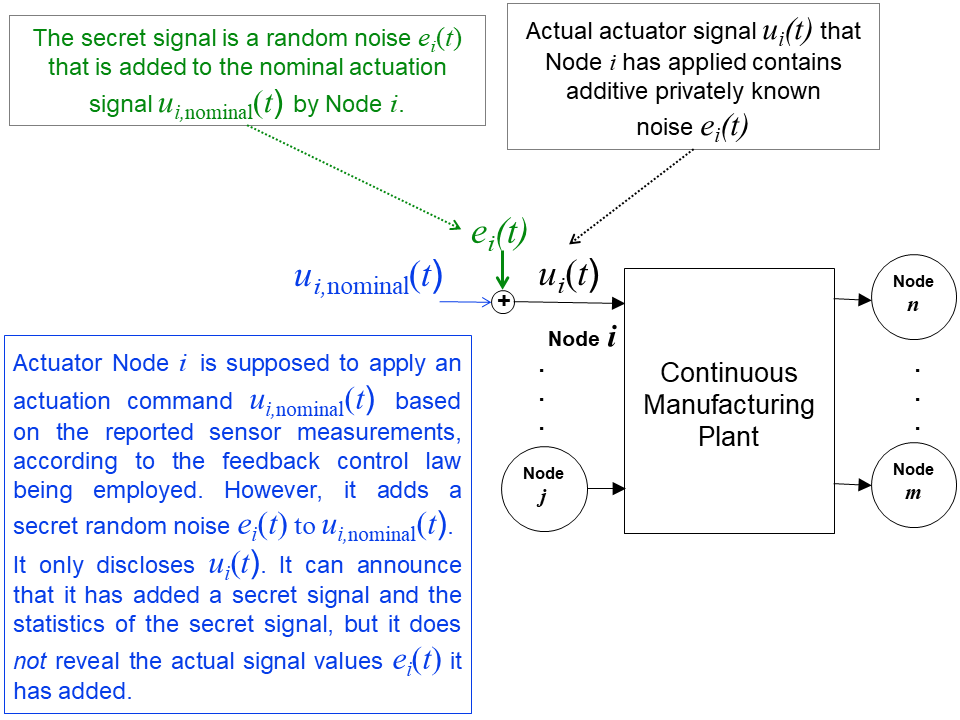}
\caption{\small Dynamic Watermarking: The Actuator Node $i$ adds a secret noise $e_i(t)$, the
``watermark," to the nominal control input $u_{i,\mbox{\scriptsize{nominal}}}(t)$ that it is expected to apply given the reported sensor measurements. It can disclose that it is adding a secret noise, and it can disclose the \emph{statistics}
of the watermark, but it does \emph{not} reveal the \emph{actual value} of the random signal $e_i(t)$.}
\label{IdeaDW}
\vspace{-0.1cm}
\end{figure}

The resulting threat model is shown in Figure \ref{fig:Malicious-Sensor}.
One  or more sensors/communication/computational nodes in the DM cyberphysical system may be compromised, as indicated in
Fig.~\ref{fig:overall-problem}. 
A compromised sensor node can report any false data at any time, as shown in Fig.~\ref{fig:Malicious-Sensor}.
We do not restrict the range of false-data attacks. With this abstraction in hand, it is possible to develop an active defense based on the idea of ``dynamic watermarking"
\cite{proceedings}. The basic idea is illustrated in Figure~\ref{IdeaDW}. Consider the problem of verifying if a sensor is being truthful in reporting its plant output measurements.  The actuation nodes superimpose a small secret random ``excitation signal'' onto their nominal actuation command.

This secret excitation can be regarded as a form of ``watermarking" in the signal domain for the dynamical (control) system and hence the name dynamic watermarking.  This excitation applied into the plant manifests itself in a transformed way in the outputs of the plant -- it is indelible just like a watermark on a sheet of paper. The manner in which it is transformed depends on the dynamics of the pathway from the actuator to the particular output.  In model-based control, design engineers have a good model of this pathway. If a sensor reports measurements that do not contain the transformed watermark, then the actuator can deduce that the sensor measurements have been compromised somewhere. One can conclude that an attack is happening and act appropriately. 

The tests to determine whether the sensor measurements contain the appropriate watermark
are statistical in nature. They rely on the fact that noise is normally present in the sensor measurements, and that the
attacker cannot separate this ambient noise from the superimposed private excitation applied
by the actuator. The statistical tests that can be conducted in various scenarios are described
in \cite{proceedings,comsnets}. To illustrate the core of the idea, consider the following example. \\
\textbf{Example:} Consider a fully-observed linear scalar Gaussian controlled dynamical system described by the equation: 
\begin{align*}
x[t+1] = a x[t] + b u[t] + w[t],
\end{align*}
where ${x}[t]$ is the state of the system and ${u}[t]$ is the control input at time $t$.  ${w}[t] \sim  \mathcal{N}(0,\sigma^{2}_{w})$ is i.i.d.  noise with a Gaussian distribution. We suppose that $a, b, \sigma^{2}_{w}$ are known to the  control system designer. 
Let $z[t]$ be the measurement \emph{reported} by the  sensor. A truthful sensor reports $z[t] \equiv x[t]$, but a malicious sensor reports
$z[t] \not\equiv x[t]$.
We assume an arbitrary history-dependent feedback control policy $g$ is in place, so that the control policy-specified input is $u_{\mbox{\scriptsize{nominal}}}[t]=g_{t}({z}^t),$
where ${z}^t :=(z[1],z[2], \ldots ,z[t])$ denotes the reported measurements up to time $t$. This results in a closed loop system, $x[t+1] = a x[t] + b u_{\mbox{\scriptsize{nominal}}}[t] + w[t].$
Suppose that the actuator superimposes a Gaussian noise  unknown to the sensor on its control input: ${u}[t]=u_{\mbox{\scriptsize{nominal}}}[t]+{e}[t],$
where  ${e}[t] \sim  \mathcal{N}(0,\sigma^{2}_{e})$ is a ``dynamic watermark.". The true state therefore satisfies:
\begin{align}
    x[t+1] - a x[t] - b u_{\mbox{\scriptsize{nominal}}}[t] & \sim & N(0, \sigma^{2}_{w}), \mbox{ and} \label{first-variance} \\
    x[t+1] - a x[t] & \sim & N(0, b^{2} \sigma^{2}_{e} + \sigma^{2}_{w}). \label{second-variance}
\end{align}
The intuition behind dynamic watermarking is that by superimposing the private excitation that is \textit{unknown} to the sensor, the actuator forces the sensor to report measurements that are correlated with $\{e[t]\}$, lest it be exposed. 
In particular, for this scalar system, the following two ``Attack Detector Tests" can be done by the actuator to detect if the sensor is malicious:\\
\emph{Attack Detector Test 1:}  Actuator checks if the reported sequence of measurements $\{z[t]\}$ satisfies\\ $\lim_{T \to \infty} \frac{1}{T}\sum_{t=0}^{T-1} ({z}[t+1]-a {z}[t]-b u_{\mbox{\scriptsize{nominal}}}[t] - b e[t])^{2} = \sigma^{2}_{w}.$
\emph{Attack Detector Test 2:}  Actuator checks if the reported sequence of measurements $\{z[t]\}$ satisfies\\ $\lim_{T \to \infty} \frac{1}{T}\sum_{t=0}^{T-1} ({z}[t+1]-a {z}[t]-b u_{\mbox{\scriptsize{nominal}}}[t])^{2} = b^{2} \sigma^{2}_{e} + \sigma^{2}_{w}.$
If the sensor is honest and reports truthful measurements $z[t] \equiv x[t]$, it passes both Tests.
If either test fails, the actuator can declare the presence of a malicious sensor in the system. 

The more difficult question is: \emph{If the signal $z[t]$ passes both tests 1 and 2, then what guarantees can we provide
on the DM CPS}? Rather strong guarantees can be provided if the signal passes both tests.
Let  ${v}[t+1] := {z}[t+1]- a{z}[t]- b u_{\mbox{\scriptsize{nominal}}}[t]- b {e}[t]-{w}[t]$. 
It has the interpretation as the \emph{additive distortion sequence introduced by the malicious sensors to the process noise present in the system}. If ${z}[t] \equiv {x}[t],$ then $v[t]\equiv 0$. \\
 \emph{Theorem 1 \cite{proceedings}:} Suppose that the reported sequence of measurements passes the two tests.  
$\lim_{T\to\infty}\frac{1}{T}\sum_{t=1}^{T} {v}^{2}[t]=0$. That is, $\{v[t]\}$ is a zero power signal. \\
It states that if the malicious sensors 
wish to remain undetected by passing the above two tests employed by the actuators, then the {\emph{only}} attack that they can launch is to distort the process noise in the system by adding a {\emph{zero power}} signal to it.
This in turn allows dynamic watermarking to provide powerful guarantees
on the overall closed-loop performance of the DM Plant even under attack.
Suppose, for example, that $|a|<1$ and a closed-loop linear control law has been designed to maintain stability, $u_{\mbox{\scriptsize{nominal}}}[t] = fx[t]$ with $|a+bf| < 1$, with the control gain $g$ chosen to yield good quadratic regulator performance. \\ 
 \emph{Theorem 2 \cite{proceedings}:}
The malicious sensor cannot compromise the mean-square performance if it is to remain undetected through the above two tests: $ \lim_{T \to \infty} \frac{1}{T} \sum_{t=0}^{T-1} x^2[t] = { (\sigma_w^2 + B^2 \sigma_e^2) }/{(1 - |a+bf|^2 )}$. \\
System metrics such as the quadratic regulation cost cannot be degraded by the malicious sensors, no matter what attack strategy they employ, without being detected. 

Dynamic watermarking is only designed to detect an attack. What is to be done after an attack is detected depends on the context. In some plants, one may be able to switch to manual control. In others, one may be able to replace the sensor, or reboot the system. Dynamic watermarking is an active defense in which the actuators inject secret excitation in order to monitor the system and detect any adversarial presence. This idea was introduced in \cite{sinopoli_replay} to detect replay attacks, and extended in \cite{physical_watermarking2} to detect other attacks. The papers \cite{proceedings, comsnets,cdc} develop detectors that provably detect arbitrary attacks that introduce non-zero power distortion. %
Dynamic Watermarking is a general methodology that can apply in a variety of contexts. It has been implemented in a laboratory process control system \cite{kim2019cyber}.
Similarly, a laboratory demonstration showing the efficacy of dynamic watermarking in an automation transportation testbed \cite{CNS} was followed by an implementation on a real autonomous vehicle driven in autonomous mode \cite{ShangguanChourKoKimKamathSatchidanandanGopalswamyKumar2020}. %
It holds potential to be deployed as a general purpose detection strategy in DM  and continuous manufacturing plants, and in IoT and manufacturing systems with sensors and actuators.

\subsection{Security of Design files: Obfuscating Designs \cite{gupta2017obfuscade}}

\begin{figure}
  \centering
  	\centering
    \includegraphics[width=\columnwidth]{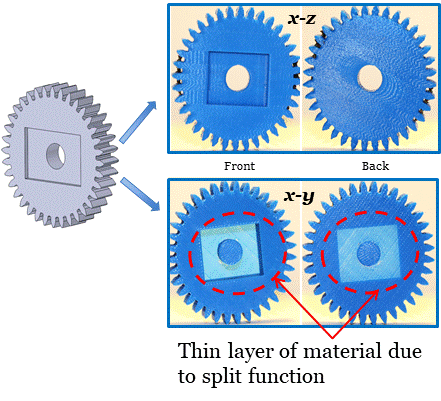}
\caption{\small The same CAD model of a gear shows different physical geometry when it is sliced and printed on the 3D printer build plate in the x-z and x-y orientations due to the security features embedded in it.}
\label{securecad}
\vspace{-0.1cm}
\end{figure}

A major concern in the DM is the security and authenticity of CAD files. These files provide incredible capabilities and information to the designers. For example, some design software programs save the entire workflow as a feature tree that the designers can use to conveniently recall a previous design step by a single click. Such capabilities are security risks because these files reveal not only the design but also the design process. Hence, embedding security in the design files may compromise some of the functionalities \cite{CHEN2017182}. 

Recent studies have shown the possibility of embedding a layer of security in the form of design features. These features can be developed with design elements such as overlapping surfaces, curvatures, and scaling functions. A part 3D printed from the design file containing such security features will appear to be different than the onscreen representation of the geometry unless the security key is applied. An example of such secure CAD file is shown in Figure~\ref{securecad}, where a stolen CAD file will print with a different gear geometry if the file is not sliced and printed in the prescribed orientation. A combination of slicing orientation, slicing resolution, printer resolution and other manufacture-time processing parameters can be used for designing such security features. 

\subsection{Securing Manufactured Parts by  Embedding Codes\cite{chen2019embedding}}

Parts manufactured by subtractive or formative manufacturing rely on surface markings for identification or authentication. Serial number, bar code, QR codes, and identifications are stamped or embossed on the parts. Additive manufacturing presents a unique possibility of encoding information in the part during manufacturing because the part is printed layer by layer. Either conventional or bespoke identification marks can be encoded in the product. These internal markings can be read by imaging methods such as tomography, radiography, and ultrasonic imaging. We have demonstrated embedding a QR code inside the part~\cite{CHEN2017182}. The method of embedding the internal identification codes depends on the AM technology. For example, sintering temperature can be changed locally to generate a feature that provides a different signature when the product is subjected to tomography. Methods such as selective laser sintering have a resolution of only a few microns so an individual feature of such size is not a concern in terms of the mechanical properties of the part. The method demonstrated slices a larger QR code into hundreds of pixel sized parts. These parts are spatially distributed in a large number of slices of the part after the slicing operation. Each part is below the critical size to compromise the mechanical properties. Slicing the code into hundreds of parts makes it difficult to find the unique direction from which it becomes a scannable code. Such obfuscations can be designed  to work in a number of ways. For instance, the sliced codes can be oriented such that the code is present in the CAD/STL files but slicing will remove it and produce a solid part without a trace of the code. 

\begin{figure}[t!]
  \centering
  	\centering
    \includegraphics[width=\columnwidth]{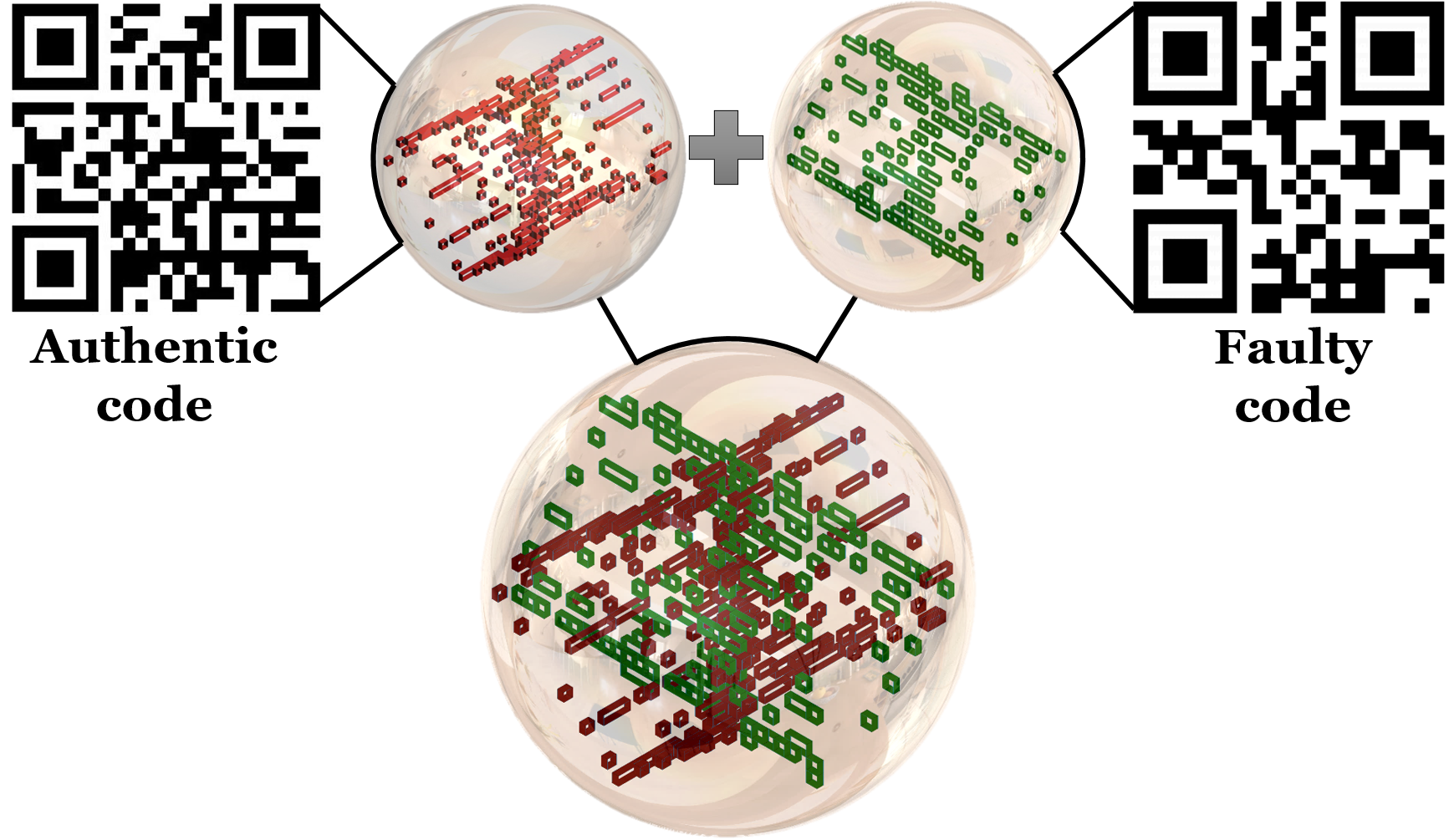}
\caption{\small Two QR codes are sliced into 300 parts each and embedded as interpenetrating codes. The correct slicing will retain only the authentic code. Incorrect slicing will retain points that will not produce any scannable code.}
\label{obfuscode}
\vspace{-0.1cm}
\end{figure} 

Reverse engineered and reconstructed CAD files will not have the code. Hence, the parts manufactured from these files will also not have the codes. Further, the parts printed from stolen CAD files will have the code and will allow identifying the unauthorized counterfeit. In another embodiment, two inter-penetrating codes can be designed such that slicing at certain angles will remove one code with the remaining code used for identification as shown in Figure~\ref{obfuscode} \cite{CHEN2017182}. This scheme will result in reverse engineered CAD files that do not resemble the original ones.

\subsection{IP Protection by Fingerprinting in Acoustic Domain \cite{behera2019system}}

\begin{figure}[b!]
  \centering
  	\centering
    \includegraphics[width=\columnwidth]{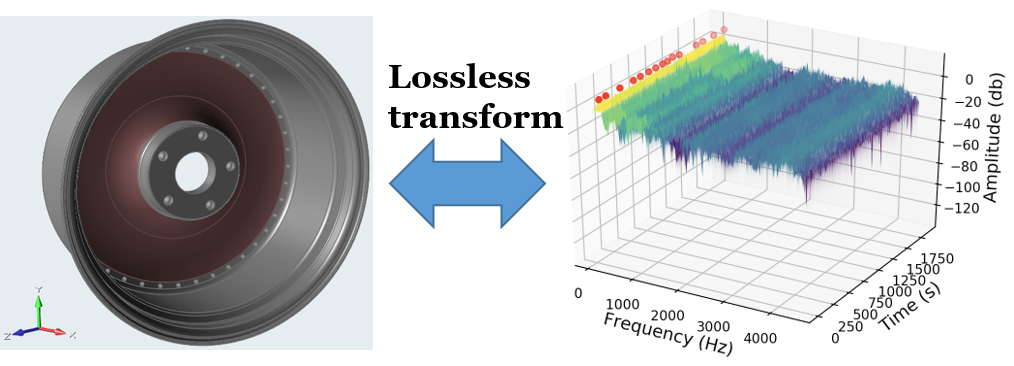}
\caption{\small Lossless transformation of a wheel hub solid model from a CAD format to a frequency domain spectrogram.}
\label{audiocad}
\vspace{-0.1cm}
\end{figure} 

CAD files are inputs for 3D printers in AM. These files are not designed just for visualization of the part design but also to manufacture the part. This places limits on encryption and compression methods that can be applied to such files. Any algorithm that causes a loss of information is not useful for such application; only lossless methods are required. 

Behera et. al. \cite{behera2019system} propose a novel encryption method where a lossless algorithm converts the CAD files to frequency domain audio files. The frequency domain files are saved as spectrograms, and used to generate fingerprints of the design in the form of (time, frequency) pairs for the amplitude peaks. These fingerprints can be used as an alternate modality for file authentication in the manufacturing process chain.

Figure~\ref{audiocad} shows a CAD model of a wheel hub, which is transformed into a frequency domain spectrogram. The red dots in the spectrogram mark the fingerprints identified for the model. The number of fingerprints depend on a designer specified threshold or automatically determined based on the security level. If the spectrogram is saved or the threshold level is low enough, the spectrogram can be converted back to the CAD model without any distortion or loss of geometry. Such spectrograms are sensitive to change in the design file. Even changing a dimension to the limit of resolution of the CAD file will create detectable perturbations in the fingerprints. 

\subsection{Securing Manufacturing IoT Networks by Device Population Diversity}

\begin{figure*}[t!]
     \subfloat[\label{fig:devices1}]{%
       \includegraphics[width=0.25\textwidth]{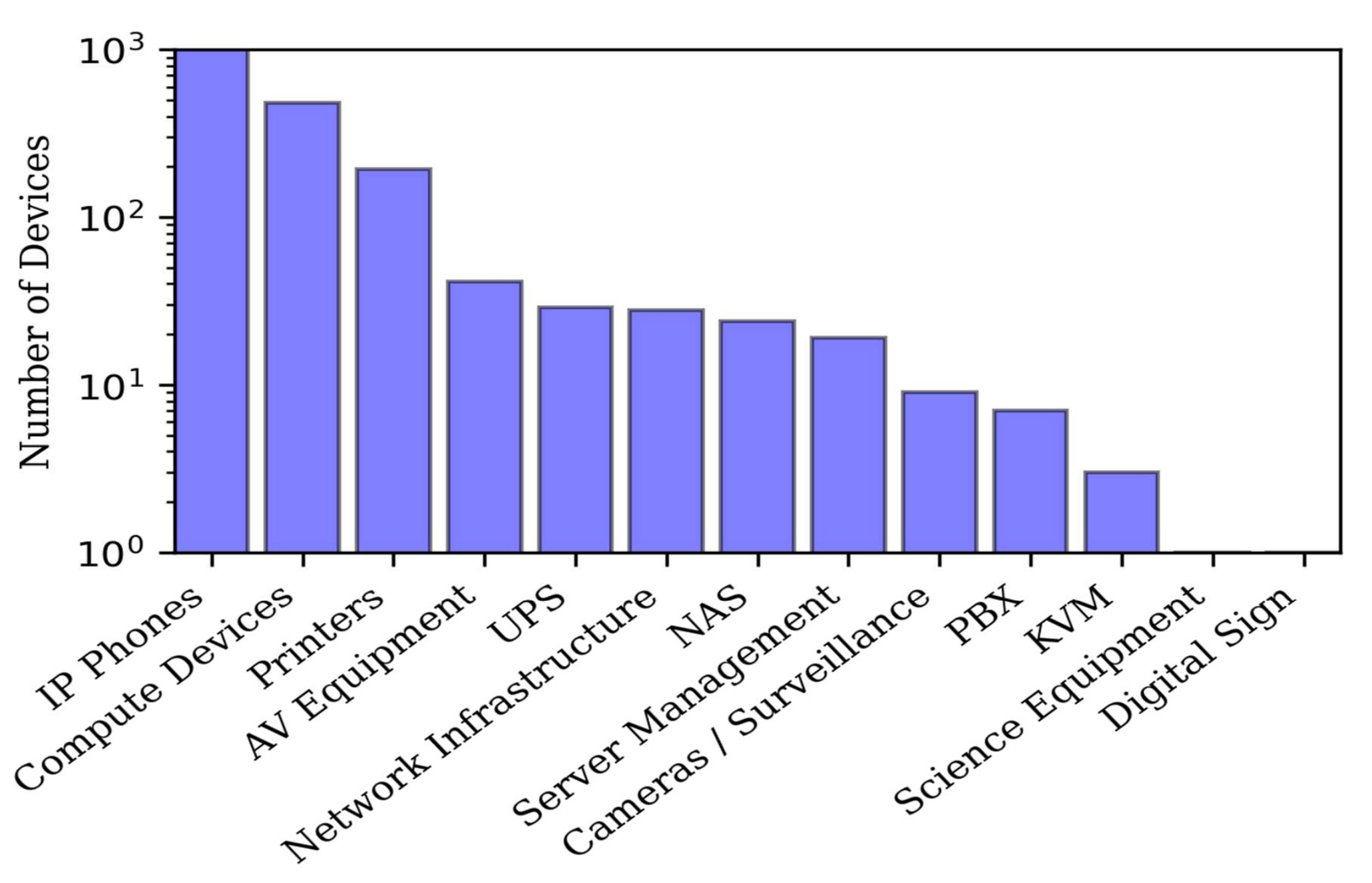}
     }
     \hspace{-0.15in}
     \subfloat[\label{fig:printers}]{%
       \includegraphics[width=0.38\textwidth]{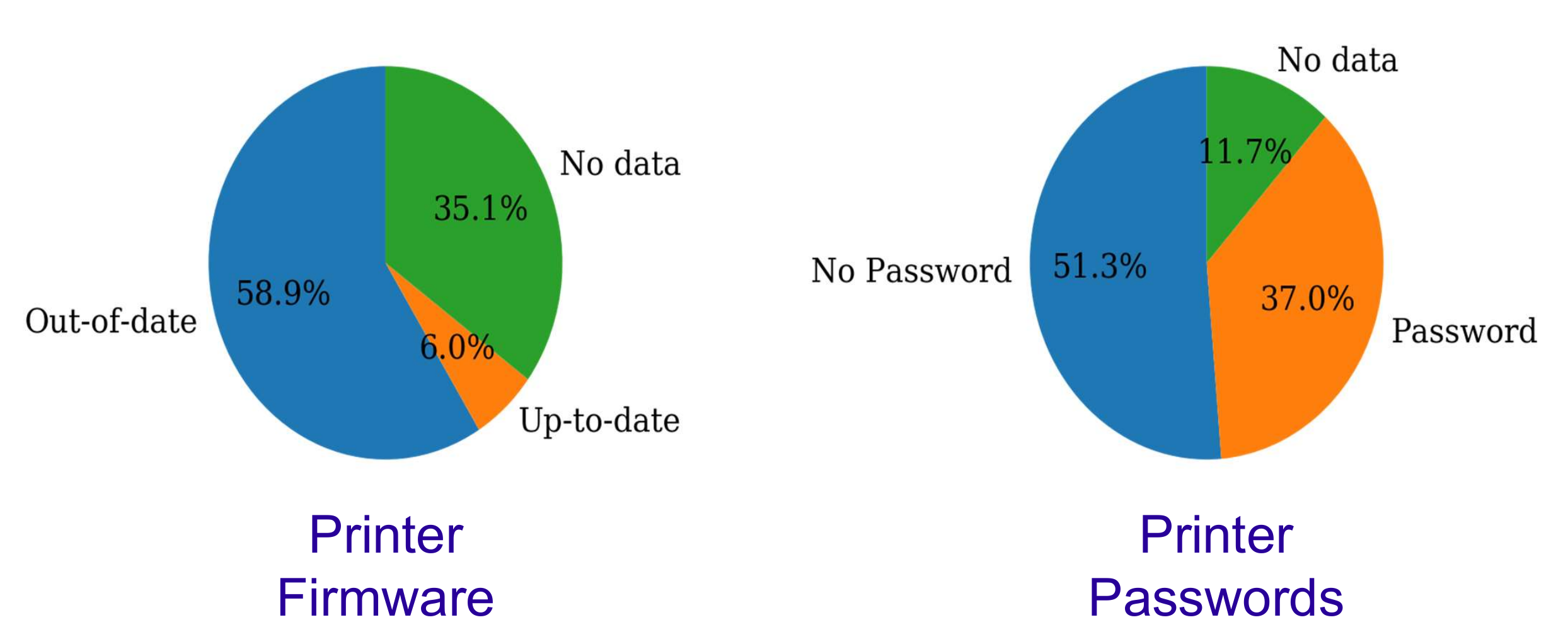}
     }
      \hspace{-0.2in}
     \subfloat[\label{fig:printer-fw}]{
     \includegraphics[width=0.38\textwidth]{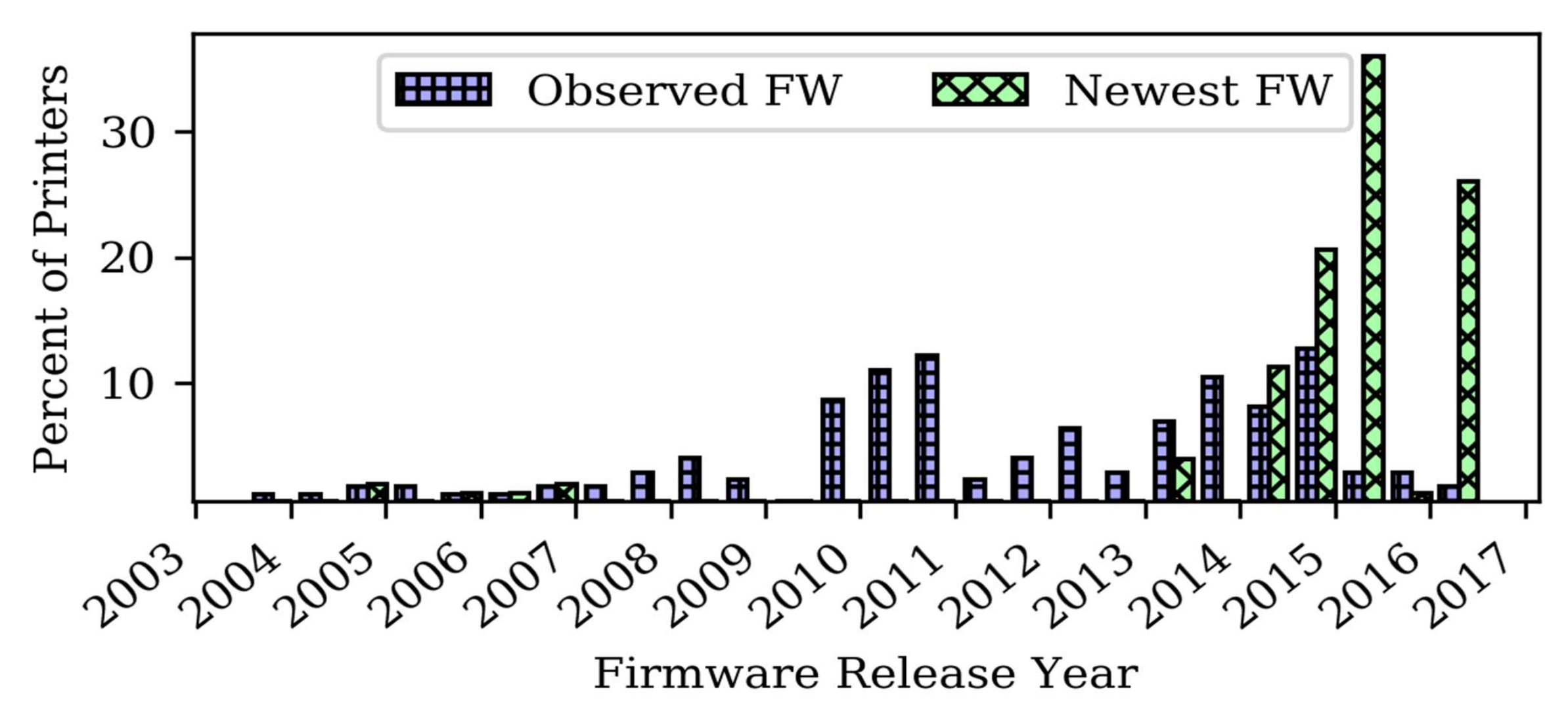}
     }
     \caption{(a) Diversity in device population on a Network. (b) Printers with no passwords (c) Status of firmware updates on printers. }
\end{figure*}

The manufacturing industry is adopting Internet-of-Things (IoT) devices at 40\% annual growth rates for enhanced asset management and increased productivity  \cite{bukkapatnamCIRP19}. The proliferation of IoT and other non-compute devices is increasing the diversity of devices connected to the network in the next-generation manufacturing system ~\cite{YangIISE2019}. The number and diversity of IoT devices is expected to grow over time as sensors and controllers are deployed widely \cite{iot-infrastructure-survey,smart-secure-iot,all-things-iot,Sivanathan2017CharacterizingAC,bac-the-model,commensurate-response,IOT-aegis}.

Due to the increasing diversity in IoT devices, their ease in connecting to networks, weak default password configurations, and general lack of ability to automatic upgrade of firmware, they are easy targets for cyberattacks
\cite{iot-security-rehash,iot-fog-survey, industrial-internet,linux-iot,consumer-iot}. While efforts to deal with vulnerability of a particular equipment or a unit in manufacturing system has been reasonably addressed, assuring cybersecurity in the presence of a diverse "population mix" of IoT sensors and other non-compute devices deployed in the next-generation manufacturing plants or across the enterprise has not received much attention.

As a proxy to studying the device population mix in a real world manufacturing enterprise, we carried out a measurement campaign of types of devices on a large-scale campus network \cite{IOT-aegis}. We carried out a census of devices connected to the campus network, and classified them based on their function. The results are shown in Figure~\ref{fig:devices1}. The devices connected to the network included desktops, laptops, mobile phones, VOIP phones, printers, TV displays, AV equipment, science appliances, and building automation gear among others. While the importance of keeping the computing equipment patched and up-to-date has for obvious reasons been recognized for quite some time, only recently the security of non-compute IoT devices is receiving attention \cite{Mirai-attack}. Our study showed that over 71\% of devices on the campus network are non-compute. Among these, $\sim$59\% of the printers on the network had out-of-date firmware (see Figure~\ref{fig:printers}) and  over half of the printers had no password. In a manufacturing plant, the percentage and diversity of non-compute devices is expected to be higher. 

Current network security approaches and tools are device agnostic and ignore the diversity of the networked IoT devices. However, not all the devices are created equal and not all the devices are updated and maintained at the same level of network hygiene. In the campus network that we studied, while the computers are managed, patched, and secured by the IT team, the printers are maintained by graduate students, the VOIP phones are managed by the communications department, and the building automation devices are maintained by the facilities department. This leads to inconsistencies in the hygiene and health across devices. We advocate enhancing security tools to consider the diversity of the device populations. \revision{As shown in Fig. \ref{fig:supply_chain}, the device population mix in a typical manufacturing floor network will look considerably different from the design network.}

Public health experts and epidemiologists consider population diversity and the differing impact of diseases on different groups in keeping the population healthy. Similarly, we advocate network security policies and mechanisms tailored to the population of devices in the manufacturing network. This has benefits over state-of-the-art device-agnostic approaches.

Dynamics of the device population has a significant impact on virus/attack epidemics in the network. For example, the Mirai attack targeted particular type of devices and networks with these devices had more compromises. Knowing the local device population allows one to mine national vulnerability database (NVD) \cite{CERT-database,nvd-database} to study vulnerabilities specific to the network. The NVD is a repository of known vulnerabilities characterized by anticipated criticality. We can construct device population specific attack vulnerability profiles.
Besides the NVD database, one could use internal information to augment the network monitoring tools. For example, a Programmable Logic Controller (PLC) controlling a boiler may need to be more carefully monitored and protected compared to a printer on the network. If additional information about the devices is available, this can be factored into allocation decisions on monitoring devices. Data from our study on campus devices revealed that the firmware in printers is not upgraded as frequently as in other devices (see Fig. \ref{fig:printer-fw}). While this knowledge is beneficial in deploying IT resources for updating/patching the device firmware to reduce the number of un-patched vulnerabilities, until that time these devices\footnote{ e.g., devices with older firmware or vulnerabilities from CERT database.} are upgraded, extra resources maybe needed to monitor them.

It is important to study the vulnerabilities of the network device population and take steps to protect local device populations. Following are at least three ways.
\begin{enumerate}
\item  Based on the number of local devices and the known vulnerabilities on these devices, network monitoring tools and resources can be optimally apportioned to maximize their effectiveness in detecting and containing the attacks. At the time of connection, the level of provided network service can be tailored to the known security vulnerabilities of the device requesting network service. The levels of service could include complete denial of service, limited access through security perimeters, requiring security patches or upgrades before providing full access to the network. These approaches apply to one device at a time at the time of connecting to the network.
\item Isolate similarly vulnerable devices on a Virtual LAN (VLAN) to provide suitable security for these devices. For example, the Windows8 devices for which no new security patches will be available could be isolated in a separate VLAN and protect them with a security device that carefully monitors Windows8 specific attacks. Similarly, IoT devices in a critical infrastructure could be put on a separate VLAN that only trusted users can access. Even if they are not perfect, such population specific isolation and protections will improve security.
\item Given the device population, network monitoring tools can aggregate anomalies based on device types to find patterns of attacks on specific types of devices. More information can be gleaned by aggregation based on device type. Observed anomalies can be checked against vulnerabilities in the NVD database to find  attack vectors.
\end{enumerate}

\section{Conclusion}
\label{sec:conclusion}

\begin{figure}[t!]
    \centering
    \includegraphics[width=1.0\columnwidth]{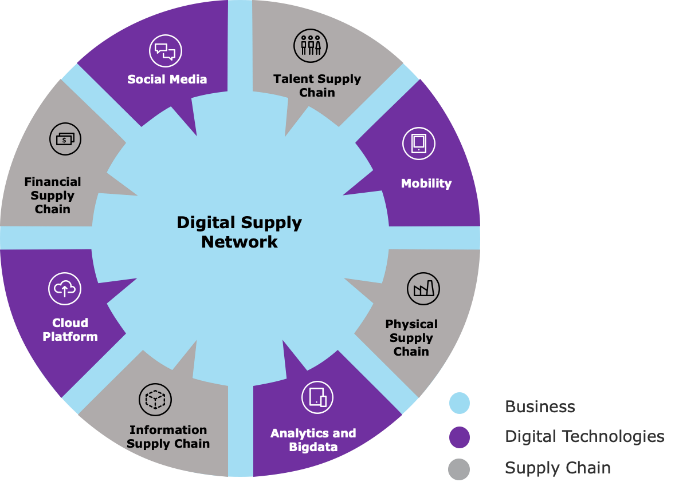}
    \caption{The emerging digital supply chain network.}
    \label{fig:my_dsn}
\end{figure}

Adoption of DM requires companies to migrate to a Digital Supply Chain Network (DSN) as shown in Figure \ref{fig:my_dsn}. The figure shows how a classical linear manufacturing supply chain collapses into a set of dynamic networks due to digitalization. DSNs enabled by networking within and across organizations are integral to the DM. While integration of the social media may be a counter-intuitive component in the DSN, companies are adopting social media platforms to report service outages and system malfunctions and for customer support. As our study shows, the elements of the DM process chain open up large attack surface and introduce many vulnerabilities making them susceptible to traditional cyberattacks and attacks that impact the physical DM and quality of manufactured products. Digitalization of the entire DM supply chain while making the production and movement of goods efficient, increases the attack surface and introduces new attack vectors. 

Not all participants in a manufacturing supply chain may have the same level of resources to implement the most advanced defenses.  The weakest links in a supply chain may besides compromising their own assets, may compromise the assets of all participants in the supply chain. This is especially true for the medium scale enterprises (MSEs), with limited resources, who nevertheless have to embrace adoption of digitalization and DM. When the MSEs employ the digital thread while setting up the DM workflow and use the DSN to establish connectivity within their enterprise and across enterprises in the supply chain, they have to tackle the threats on multiple levels. The challenge for these MSEs is therefore to be judicious in using the limited resources to address these threats. The MSEs must prioritize which cybersecurity issues to address as they transition to a DM workflow.

While this study focused on cybersecurity of manufacturing-unique elements of a DSN, other elements in the DSN such as the information, financial, and business networks are also important. Some of them can be secured using well-known information security approaches such as encrypting data and authenticating the communications.  Side channel attacks and reverse engineering of products are threats that extend beyond the DM network and impact a company significantly. Reverse engineering of a product can lead to revenue loss, where the CAD models may be generated by skillful designers based on an actual part acquired from the OEM without any disruption or breaches to the connected supply chain. These additional risks need to be addressed when securing DM. Most DM IOT technology components lack sufficient device activity logging capability. Insecure network protocols are typically used to connect DM  components to the internet. Various methods can be used to assess the security posture of a manufactured product. Traditional systems have typically either been designed without security in mind, or with the explicit presumption that the system is isolated and so not subject to cyberattacks~\cite{tuptuk2018security}. 
The new generation of manufacturing sectors resulting from the adoption of the DM process workflow and migration to the DSN need special focus on securing the complex systems that are integrated within the control network in the manufacturing plant. Hence, security controls should be designed from the inception of software development to hardware configuration in the control network.

\section*{Acknowledgment}

The NYU team acknowledges the National Science Foundation Cyber-Physical Systems grant CMMI-1932264 and NSF grant DGE-1931724. R. Karri and N. Gupta are supported in part by NYU Center for Cybersecurity, and R. Karri is also supported in part by NYU-AD Center for Cybersecurity. Bukkapatnam's research is partially supported by the Natioanal Science Foundation grants CMMI-1432914 and S\&AS INT-1849085, and Texas A\&M University's x-grants program. Reddy’s research is supported by Qatar National Research Foundation grant 9-069-1-018. The material by Kumar is based upon work partially supported by NSF Science \& Technology Center Grant CCF-0939370, NSF CCF-1934904, the U.S. Army Research Office under Contract No. W911NF-18-10331, the U.S. Army Research Laboratory under Cooperative Agreement Number W911NF-19-2-0243, the U.S. Army Research Laboratory under Contract No. W911NF-19- 2-0033, U.S. ONR under Contract No. N00014-18-1-2048, and the Department of Energy under Contract No. DE-EE0009031. The views and conclusions contained in this document are those of the authors and should not be interpreted as representing the official policies, either expressed or implied, of the National Science Foundation, Army Research Office, Army Research Lab, the Office of Naval Research, the Department of Energy, or the U.S. Government. The U.S. Government is authorized to reproduce and distribute reprints for Government purposes notwithstanding any copyright notation herein.

\ifCLASSOPTIONcaptionsoff
  \newpage
\fi

\bibliographystyle{IEEEtran}
\bibliography{references.bib,references-reddy.bib}

\begin{IEEEbiography}[{\includegraphics[width=1in,height=1.25in,clip,keepaspectratio]{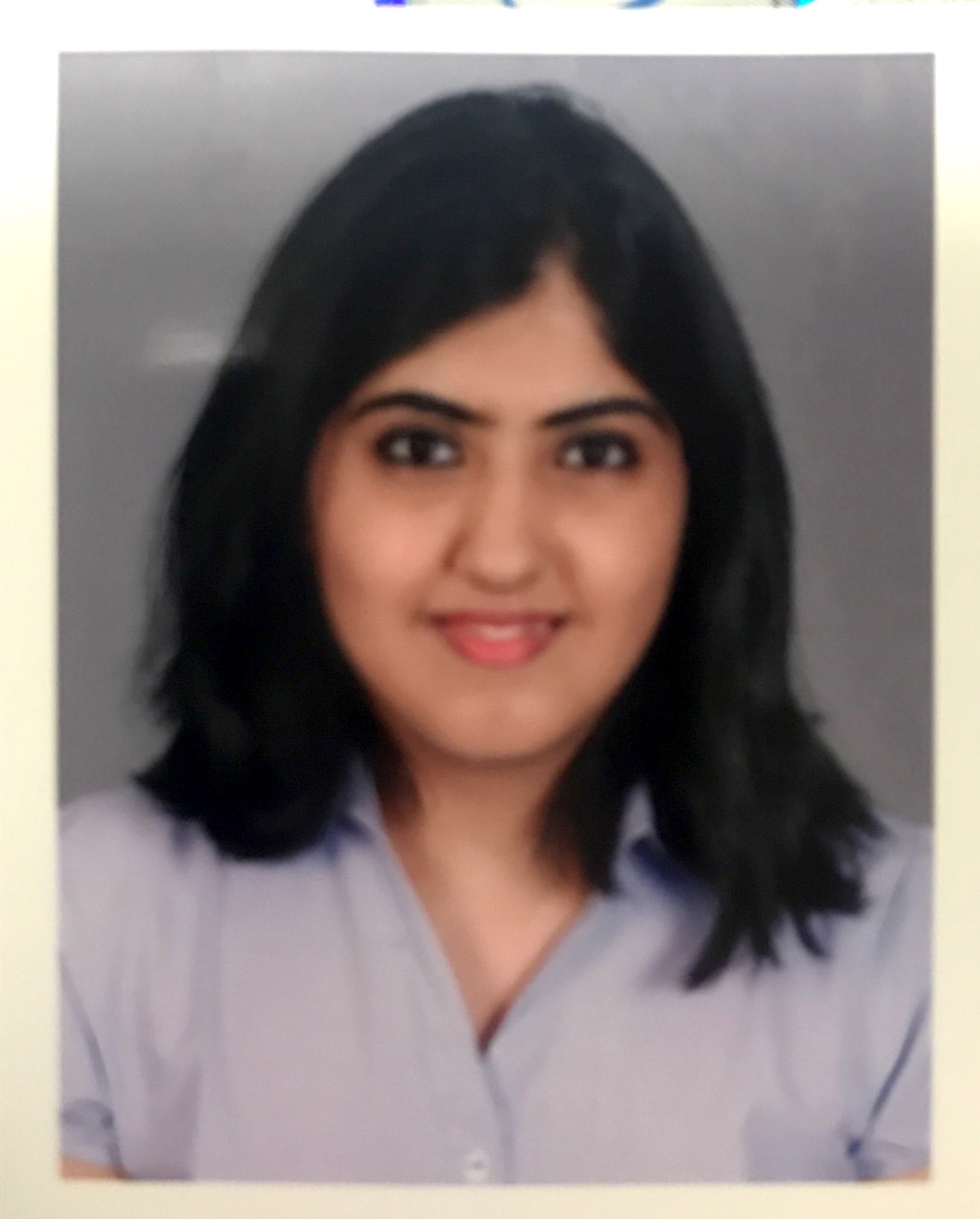}}]{Priyanka Mahesh}is a Graduate student at New York University. She obtained her B.Tech degree in Computer Science from SRM University. She has worked in the consulting industry in the field of cybersecurity on projects related to telematics and ICS security. Her research is focused on addressing security concerns in cyber-physical systems, embedded systems and industrial control systems in order to build trustworthy IOT systems.
\end{IEEEbiography}

\begin{IEEEbiography}[{\includegraphics[width=1in,height=1.25in,clip,keepaspectratio]{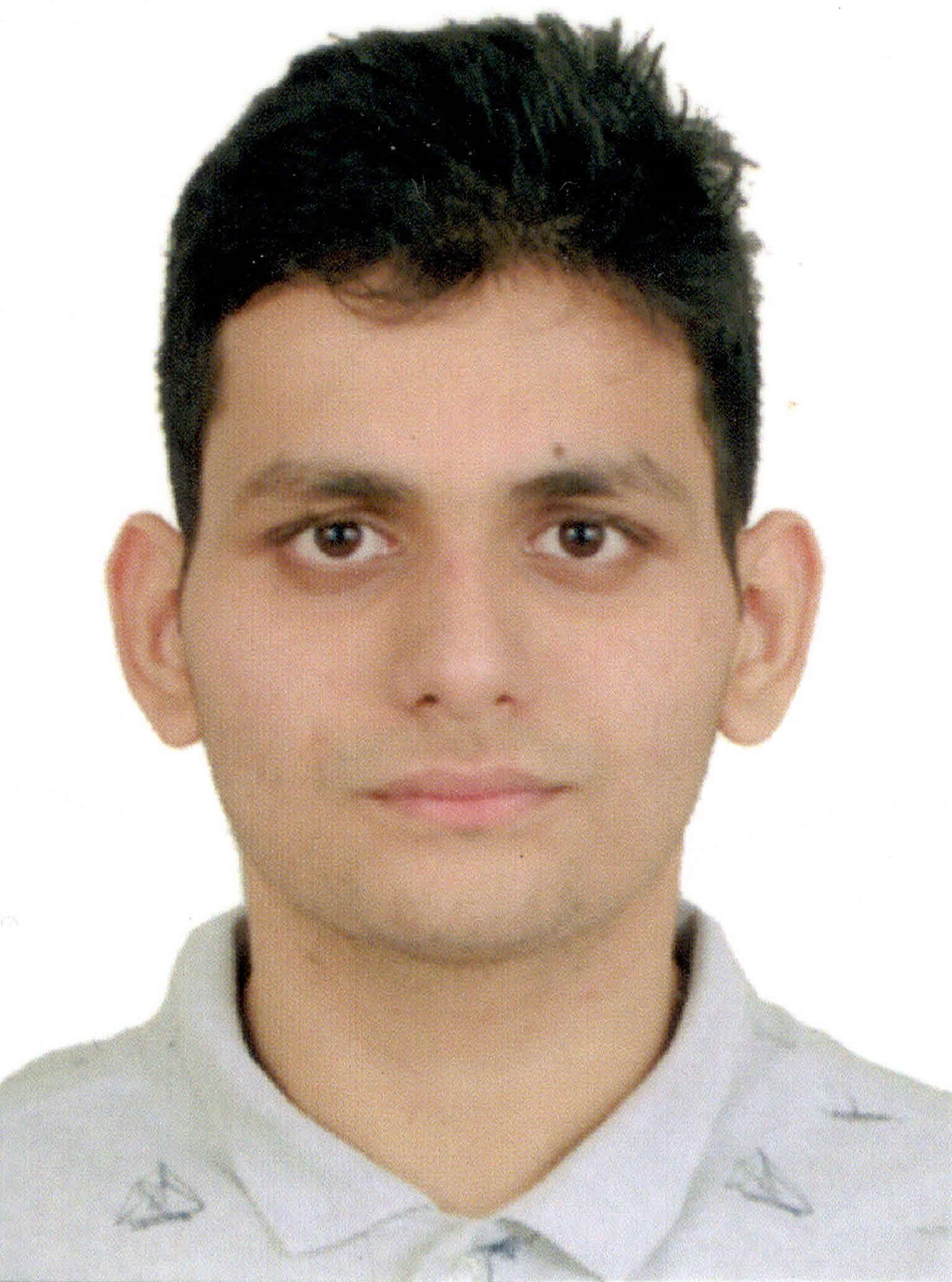}}]{Akash Tiwari}received the B.Tech. degree in industrial and systems engineering from the Indian
Institute of Technology (IIT) Kharagpur, India, in 2019. He is currently pursuing the Ph.D. degree with the Department of Industrial and Systems Engineering, Texas A\&M University, College Station,
TX, USA. He was a Summer Intern with the Royal Enfield
Motors Factory, Chennai, India, in 2017. In 2018, he was a Summer Research Intern with the Durham Univeristy Business School, Durham, U.K. 
\end{IEEEbiography}

\begin{IEEEbiography}[{\includegraphics[width=1in,height=1.25in,clip,keepaspectratio]{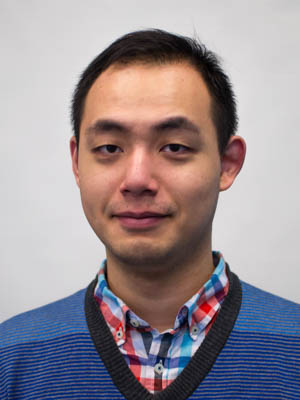}}]{Chenglu Jin} is joining CWI Amsterdam as a tenure-track researcher. He was a research assistant professor at NYU Center for Cybersecurity and Center for Urban Science and Progress. His research interest is cyber-physical system security, hardware security, and applied cryptography. He holds a Ph.D. degree from the University of Connecticut, Electrical and Computer Engineering Department. He has published papers in major conferences/journals such as TCHES, AsiaCCS, ACSAC, FC, HOST, TDSC, etc.
\end{IEEEbiography}

\begin{IEEEbiography}[{\includegraphics[width=1.0in,height=1.25in,clip,keepaspectratio]{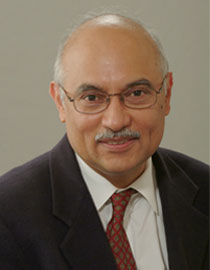}}]{P. R. Kumar} (F’88) received the B.Tech. degree in electronics engineering from Indian Institute of Technology (IIT) Madras, Chennai, India, in 1973, and the D.Sc. degree in systems science and mathematics from Washington University in St. Louis, St. Louis, MO, USA, in 1977. He is currently with Texas A\&M University, College Station, TX, USA. He was a faculty member with the University of Maryland, Baltimore County (1977–1984) and the University of Illinois at Urbana-Champaign (1985–2011). He was the Leader of the Guest Chair Professor Group on Wireless Communication and Networking with Tsinghua University. He is an Honorary Professor with IIT Hyderabad. His research interests include cyber-physical systems, cybersecurity, privacy, wireless networks, renewable energy, smart grid, autonomous vehicles, and unmanned air vehicle systems. Prof. Kumar is a member of the U.S. National Academy of Engineering, The World Academy of Sciences, and the Indian National Academy of Engineering. He was awarded a Doctor Honoris Causa by ETH Zurich. He was the recipient of the IEEE Field Award for Control Systems, the Donald P. Eckman Award of the AACC, Fred W. Ellersick Prize of the IEEE Communications Society, the Outstanding Contribution Award of ACM SIGMOBILE, the INFOCOM Achievement Award, and the SIGMOBILE Test-of-Time Paper Award. He is a Fellow ACM. He was also the recipient of the Distinguished Alumnus Award from IIT Madras, the Alumni Achievement Award from Washington University in St. Louis, and the Daniel Drucker Eminent Faculty Award from the College of Engineering, University of Illinois at Urbana-Champaign.
\end{IEEEbiography}

\begin{IEEEbiography}[{\includegraphics[width=1in,height=1.25in,clip,keepaspectratio]{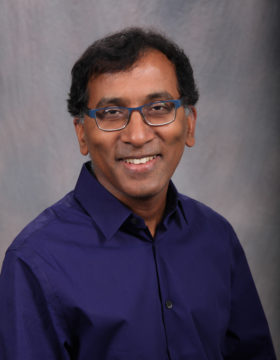}}]{Narasimha Reddy}
is currently a J.W. Runyon Professor in the department of Electrical and Computer Engineering at Texas A\&M University as well as the Associate Dean for Research with the Texas A\&M Engineering Program and the Assistant Director of Strategic Initiatives \& Centers with the Texas A\&M Engineering Experiment Station. Reddy’s research interests are in Computer Networks, Storage Systems, and Computer Architecture. During 1990-1995, he was a Research Staff Member at IBM Almaden Research Center in San Jose. Reddy holds five patents and was awarded a technical accomplishment award while at IBM. He received an NSF Career Award in 1996. His honors include an Outstanding Professor award by the IEEE student branch at Texas A\&M during 1997-1998, an Outstanding Faculty award by the Department of Electrical and Computer Engineering during 2003-2004, a Distinguished Achievement award for teaching from the Former Students Association of Texas A\&M University, and a citation “for one of the most influential papers from the 1st ACM Multimedia Conference”.
\end{IEEEbiography}

\begin{IEEEbiography}[{\includegraphics[width=1in,height=1.25in,clip,keepaspectratio]{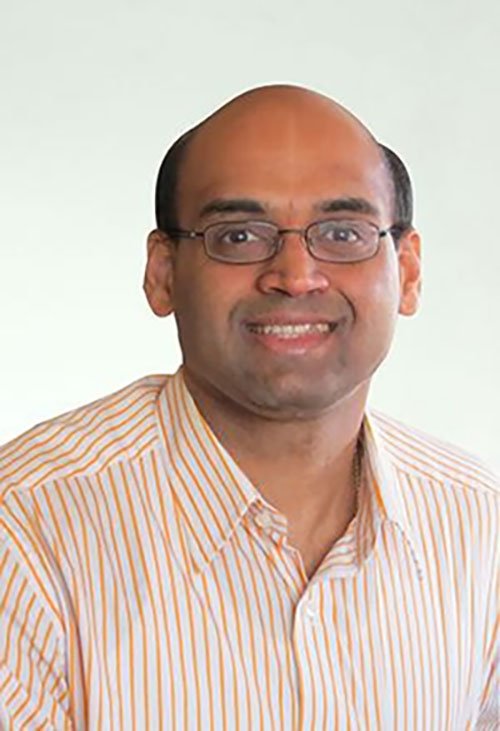}}]{Satish T.S. Bukkapatnam}
received his Ph.D. and M.S. degrees in industrial and manufacturing engineering from the Pennsylvania State University. He currently serves as Rockwell International Professor with the Department of Industrial and Systems Engineering department at Texas A\&M University, College Station, TX, USA, and has been selected as a Fulbright-Tocqueville distinguished chair. He is also the Director of Texas A\&M Engineering Experimentation Station (TEES) Institute for Manufacturing Systems. His research in smart manufacturing addresses the harnessing of high-resolution nonlinear dynamic information, especially from wireless MEMS sensors, to improve the monitoring and prognostics, mainly of ultra-precision and nano-manufacturing processes and machines, and wearable sensors for cardio-respiratory processes. His research has led to over 160 articles in journals and conference proceedings. He is a fellow of the Institute for Industrial and Systems Engineers (IISE), and the Society of Manufacturing Engineers (SME).
\end{IEEEbiography}

\begin{IEEEbiography}[{\includegraphics[width=1in,height=1.25in,clip,keepaspectratio]{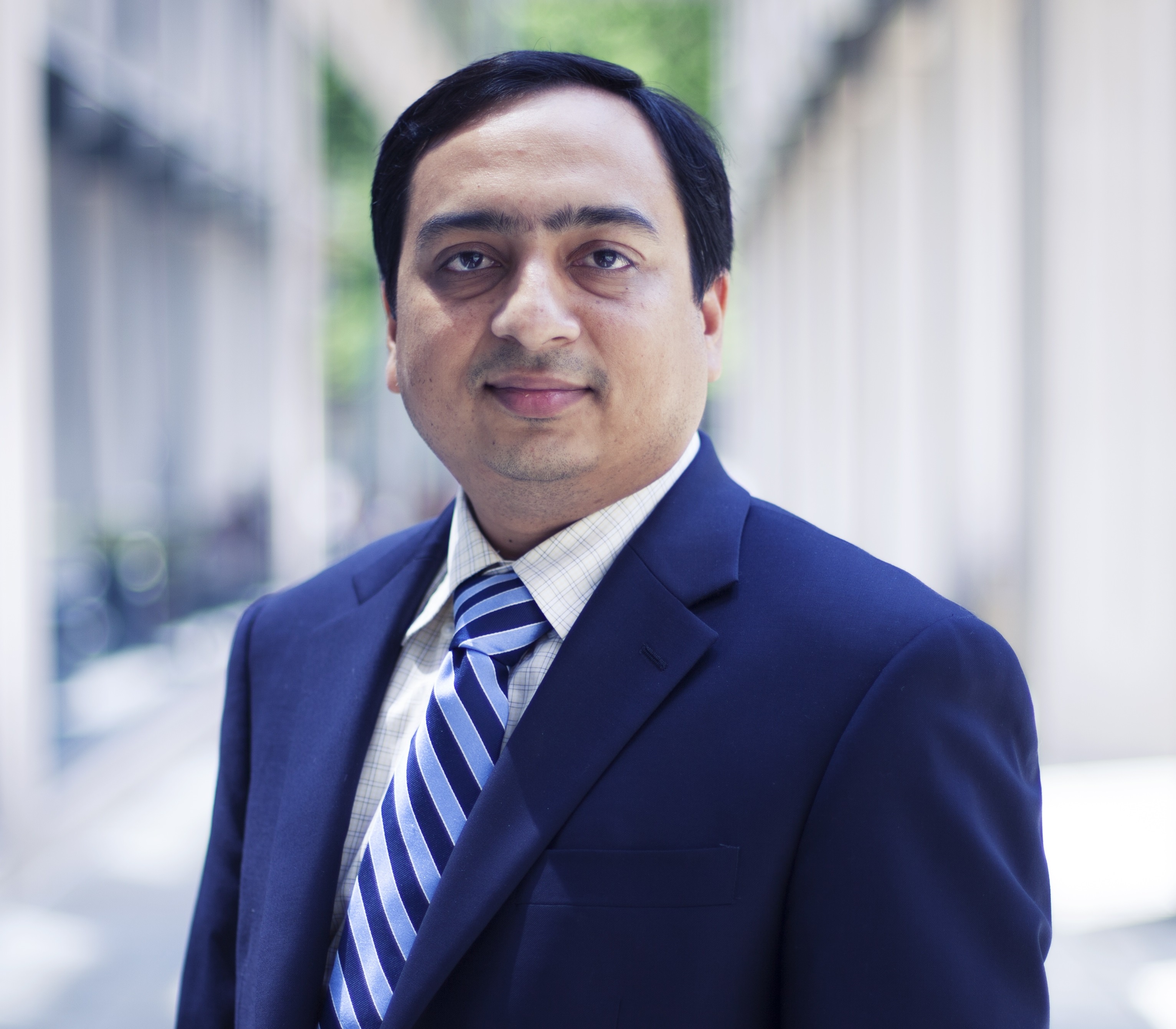}}]{Nikhil Gupta}
is a Professor of Mechanical and Aerospace Engineering at New York University. He is also affiliated with NYU Center for Cybersecurity. His research is focused on developing methods to secure computer aided design files against theft of intellectual property and unauthorized production of parts. His group is also using machine learning methods for reverse engineering of parts and mechanical property characterization. He is an author of over 195 journal articles and book chapters on composite materials, materials characterization methods and additive manufacturing security.
\end{IEEEbiography}

\begin{IEEEbiography}[{\includegraphics[width=1.0in,height=1.25in,clip,keepaspectratio]{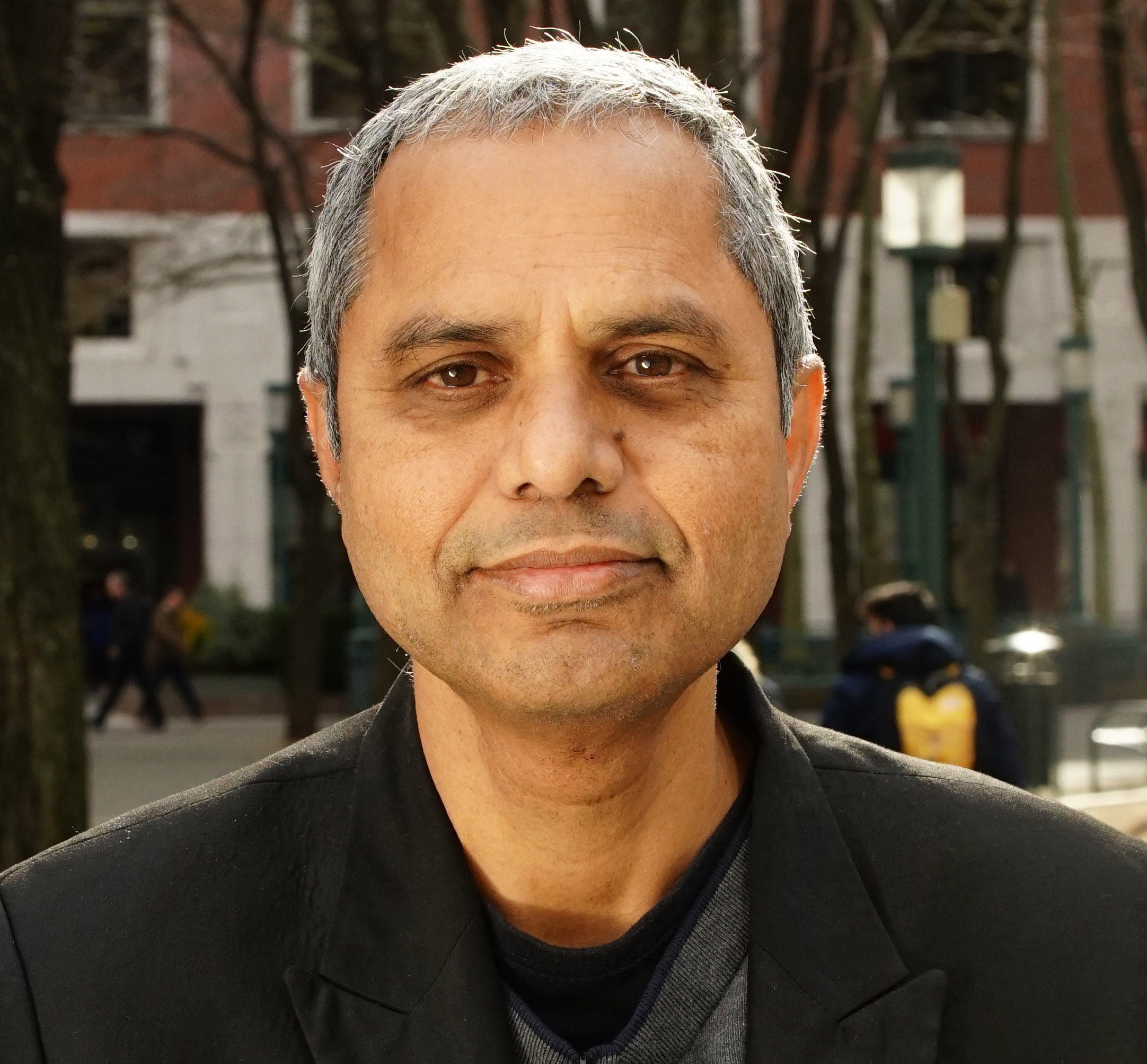}}]{Ramesh Karri} (F'20) 
 is a Professor of Electrical and Computer Engineering at New York University. He co-directs the NYU Center for Cyber Security (http://cyber.nyu.edu). He co-founded the Trust-Hub (http://trust-hub.org) and organizes the Embedded Systems Challenge (https://csaw.engineering.nyu.edu/esc), the annual red team blue team event.  Ramesh Karri has a Ph.D. in Computer Science and Engineering, from the University of California at San Diego and a B.E in ECE from Andhra University. His research and education activities in hardware cybersecurity include trustworthy integrated circuits, processors and cyber-physical systems; security-aware computer-aided design, test, verification, validation, and reliability; nano meets security; hardware security competitions, benchmarks, and metrics; biochip security; additive manufacturing security. He has published over 275 articles in leading journals and conference proceedings. His work in trustworthy hardware received best paper award nominations (ICCD 2015 and DFTS 2015), awards (ACM TODAES 2017, ITC 2014, CCS 2013, DFTS 2013 and VLSI Design 2012, ACM Student Research Competition at DAC 2012, ICCAD 2013, DAC 2014, ACM Grand Finals 2013, Kaspersky Challenge and Embedded Security Challenge). He received the Humboldt Fellowship and the National Science Foundation CAREER Award.  He is a Fellow of the IEEE for his contributions to and leadership in Trustworthy Hardware. He is the Editor-in-Chief of ACM Journal of Emerging Technologies in Computing. Besides, he served/s as the Associate Editor of IEEE Transactions on Information Forensics and Security, IEEE Transactions on CAD, ACM Journal of Emerging Computing Technologies, ACM Transactions on Design Automation of Electronic Systems (2014-), IEEE Access, IEEE Transactions on Emerging Technologies in Computing, IEEE Design and Test (2015-) and IEEE Embedded Systems Letters (2016-). He served as an IEEE Computer Society Distinguished Visitor (2013-2015). He served on the Executive Committee of the IEEE/ACM Design Automation Conference leading the Security@DAC initiative (2014-2017). He has given keynotes, talks, and tutorials on Hardware Security and Trust. 
\end{IEEEbiography}

\end{document}